\documentclass{pasj00}
\draft

\usepackage{color}
\usepackage{ulem}

\begin{document}
\SetRunningHead{Y. Ezoe et al.}{
Enhancement of Terrestrial Diffuse X-ray Emission
Associated With Coronal Mass Ejection
and Geomagnetic Storm
}
\Received{}
\Accepted{}

\title{
Enhancement of Terrestrial Diffuse X-ray Emission
Associated With Coronal Mass Ejection
and Geomagnetic Storm
}

\author{
  Yuichiro \textsc{Ezoe}\altaffilmark{1},
  Yoshizumi \textsc{Miyoshi}\altaffilmark{2}, 
  Hiroshi \textsc{Yoshitake}\altaffilmark{3},
  Kazuhisa \textsc{Mitsuda}\altaffilmark{3},\\
  Naoki \textsc{Terada}\altaffilmark{4}, 
  Shihoko \textsc{Oishi}\altaffilmark{1}, 
  and Takaya \textsc{Ohashi}\altaffilmark{1} 
}

\altaffiltext{1}{
  Tokyo Metropolitan University, 1-1, Minami-Osawa, Hachioji,
  Tokyo, 192-0397, JAPAN
}

\altaffiltext{2}{
  Nagoya University, 
  Furo-cho, Chikusa-ku, Nagoya 464-8601, JAPAN
}

\altaffiltext{3}{
  The Institute of Space and Astronautical Science (ISAS),
  Japan Aerospace and eXpoloration Agency (JAXA),               
  3-1-1 Yoshinodai, Sagamihara, Kanagawa 229-8510, JAPAN
}

\altaffiltext{4}{
  Tohoku University, 
  6-3, Aoba, Aramakiaza, Sendai, Miyagi 980-8578, JAPAN
}

\email{ezoe@phys.metro-u.ac.jp}

\KeyWords{X-ray: diffuse background --- Sun: solar wind --- Sun: solar-terrestrial relations --- Earth} 

\maketitle

\begin{abstract}
We present an analysis of a Suzaku observation taken during the geomagnetic 
storm of 2005 August 23-24. We found time variation of diffuse soft 
X-ray emission when a coronal mass ejection hit Earth and caused 
a geomagnetic storm. The diffuse emission consists 
of fluorescent scattering of solar X-rays and exospheric solar
wind charge exchange. 
The former is characterized by a neutral oxygen emission line
due to strong heating of the upper atmosphere during the storm time,
while the latter is dominated by a sum of C\emissiontype{V},
C\emissiontype{VI}, N\emissiontype{VI}, N\emissiontype{VII},
O\emissiontype{VII}, and O\emissiontype{VIII} emission lines
due to the enhanced solar wind flux in the vicinity of the exosphere.   
Using the solar wind data taken with the ACE and WIND satellites,
a time correlation between the solar wind and the strong 
O\emissiontype{VII} line flux were investigated. We estimated necessary 
column densities for the solar X-ray scattering and exospheric 
SWCX. From these results, we argue that a part of the solar 
wind ions enter inside the magnetosphere and cause the SWCX reaction.
\end{abstract}

\section{Introduction}
\label{sec:intro}

A coronal mass ejection (CME) is a large burst of coronal magnetic fields
and plasma with a typical mass of 10$^{15-16}$ g and speeds
of 250-1000 km s$^{-1}$ into interplanetary space \citep{gos97,hud10}. 
CMEs are often associated with solar flares and prominence eruptions. 
The occurrence of CMEs depends on the phase of the solar cycle.
The ejected CMEs that move toward Earth drive interplanetary shocks 
and trigger geomagnetic storms (e.g., \cite{gon94,miy05}). 
Such geomagnetic storms pose significant hazards to space operations.
Enhancements of trapped particles of the radiation belts increase 
spacecraft charging and solar energetic protons cause single 
event upset \citep{pil06}.
Atmospheric heating by the charged particles 
and solar ultraviolet/X-ray emission causes the Earth's upper 
atmosphere to expand, and leads to satellite drag \citep{doo06}.

For general users of X-ray astronomy satellites, signals associated 
with CMEs and geomagnetic storms may constitute additional sources of 
background when they observe astrophysical objects. 
Increased scattering of solar X-rays 
by the Earth's atmosphere is 
often seen in satellite data.
This scattering is due to Thomson scattering of solar X-rays by electrons in the 
sunlit atmosphere and absorption of incident solar 
X-rays followed by the emission of characteristic K lines \citep{pet00}. 
Removal of time durations when the line of sight direction is near the 
sunlit atmosphere is effective in removing this emission.

Another major background is solar wind charge exchange (SWCX).
This occurs when an ion in the solar wind interacts with a neutral atom. 
The ion strips an electron(s) from the atom, and then X-ray or ultraviolet 
photon(s) are released as the electron relaxes into the ground state.
Short-term variations from SWCX in Earth's exosphere \citep{sno04,war04,fuj07,car08,ezo10,ezo11,car11}
or longer-term variations from SWCX in interplanetary space \citep{smi05}
and heliosphere \citep{cra00,kou07,kou09}, can produce diffuse X-ray
emission below $\sim2$ keV.
To characterize the SWCX emission in an astrophysical data set, 
careful checks of an X-ray light curve and simultaneously observed
solar wind data are indispensable.

The terrestrial diffuse X-ray emission, i.e., scattering of solar X-rays
and exospheric SWCX, provides valuable information concerning the 
atmospheric expansion, the exospheric density, the constituents of 
the solar wind, and the transport processes of the plasma within the 
bow shock. 
\citet{car08} systematically searched for the exospheric SWCX emission
from data in the XMM-Newton Archive. Approximately 3.4 \% of observations were
affected by the exospheric SWCX \citep{car11}.
Most of the SWCX emission were seen when XMM-Newton observes
through the sub-solar side of the magnetosheath. 
In some cases, the SWCX emission was seen when the line of sight 
direction did not intersect the high flux region. They reasoned
that these cases probably originate from CMEs. 
They studied the most spectrally rich example
of SWCX and argued that this event was associated with a CME 
recorded on 2001 October 21 \citep{car10}.

Thanks to a good energy response and the low instrument background 
of the X-ray CCDs (X-ray Imaging Spectrometer, XIS: \cite{koy07}), 
the fifth Japanese X-ray astronomical satellite Suzaku \citep{mit07} 
is ideal for studying the terrestrial diffuse X-ray emission.
\citet{fuj07} discovered an exospheric SWCX event in the direction
of the north ecliptic pole.
Their data indicated that a distance to the point where 
the geomagnetic field becomes open to space for the
first time in the line of sight may be closely 
related to the short term SWCX variability.
\citet{ezo10} found another exospheric SWCX event toward the 
sub-solar side of the magnetosheath. They conducted a cross
correlation analysis using the Suzaku O\emissiontype{VII} 
X-ray light curve and the ACE solar wind O$^{7+}$ curve, 
and found a significant correlation between them.
The necessary column density of neutral hydrogen atoms in 
the Earth's exosphere to explain the observed X-ray flux 
exceeded that predicted by the exosphere model by a factor of 10.
They reasoned that the discrepancy can be due to uncertainty 
of the model itself and/or solar wind distribution in the 
magnetosphere.

In this paper, we present evidence of a strong enhancement of the terrestrial 
diffuse X-ray emission associated with a strong geomagnetic storm
recorded on 2005 August 24. The event under analysis occurred during
the solar declining phase of solar cycle 23.
We utilized the Dst index as an indicator of the geomagnetic storm.
This is an index for the world wide magnetic storm level and constructed 
by averaging the horizontal component of the geomagnetic field from 
mid-latitude and equatorial magnetometer data. Negative Dst values 
caused by the storm time ring current indicate a magnetic storm is 
in progress. The Dst index is provided by the World Data Center for 
Geomagnetism, Kyoto, Japan\footnote{http://wdc.kugi.kyoto-u.ac.jp/dstdir/}.
The storm studied in this paper is strongest with the 
Dst index of $-216$ nT that Suzaku has experienced as of 2011 April.

\section{Observation}
\label{sec:obs}

\begin{figure}[p]
  \begin{center}
    \FigureFile(\textwidth,){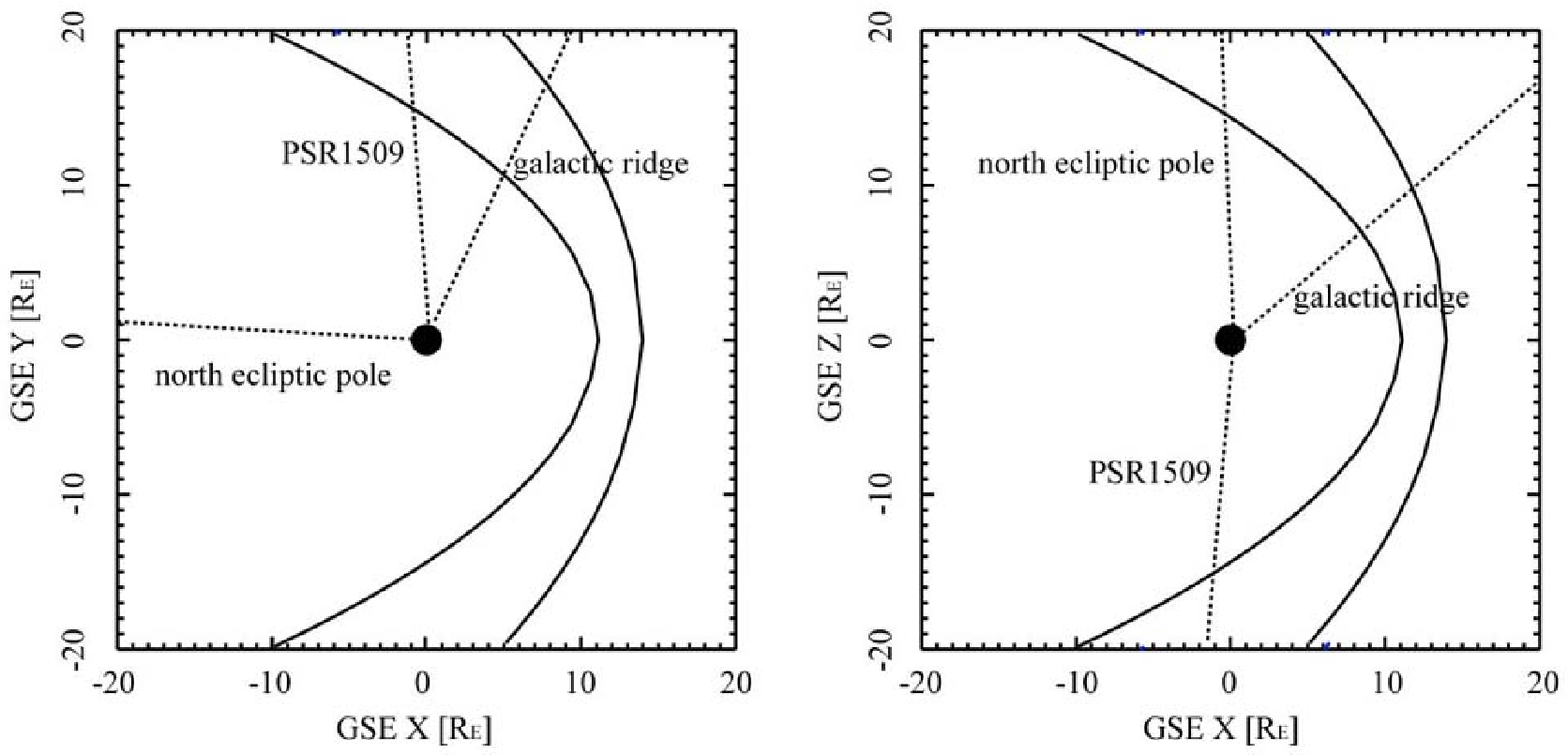}
  \end{center}
  \caption{Line of sight directions in the GSE XY and XZ planes 
    during this observation (PSR 1509),
    the north ecliptic pole \citep{fuj07}, 
    and the galactic ridge \citep{ezo10}.
    Two solid lines indicate approximate positions of 
    the magnetopause and bow shock.
  }\label{fig:los}
\end{figure}

The Suzaku observation on 2005 August 23-24 was toward the 
PSR B1509$-$58 (RA $=$ 228.484 deg, Dec $=$ $-$59.136 deg, 
$l=320.322$ deg, $b=-1.163$ deg). 
The Suzaku observation ID is 100009010.
This field contains a bright 
X-ray pulsar, a pulsar wind nebula (PWN), and the hot nebula RCW 89. 
The Galactic column in the direction of this field 
as estimated using the HEASARC nH tool\footnote{http://heasarc.gsfc.nasa.gov/cgi-bin/Tools/w3nh/w3nh.pl} 
is 1.4$\times10^{22}$ cm$^{-2}$.
The average line of sight vector in the GSE and GSM coordinates are ($-$0.0395, 
0.7717, $-$0.6343) and ($-$0.0395, 0.5811, $-$0.8124), respectively.
In figure \ref{fig:los}, we plot the line of sight during the observation.
It is toward the south direction, while the past two Suzaku detections
of exospheric SWCX events were toward the north ecliptic pole and magnetosheath.
The observation starts from 2005 August 23 (the Day of Year, DOY, of 235 in 2005) 
08:38 and ends August 24 (DOY of 236) 20:38. 

The original aim of the observation was timing calibration using the
pulsation period of the pulsar that has been published by \citet{ter08}.
For this purpose, in a part of the observation, the XIS was operated 
at timing mode, in addition to normal clocking mode. 
Because the imaging capability is not available in the timing mode, 
we analyzed only the normal clocking mode data. The effective exposure time of
the normal clocking mode data after standard data screening is 59 ks. The process version is 2.0.6.13.
The HEAsoft analysis package (version 6.4)\footnote{http://heasarc.gsfc.nasa.gov/docs/software/lheasoft/} 
was used to extract images, light curves, spectral products and instrumental response files.

\begin{figure}[p]
  \begin{center}
    \FigureFile(0.425\textwidth,){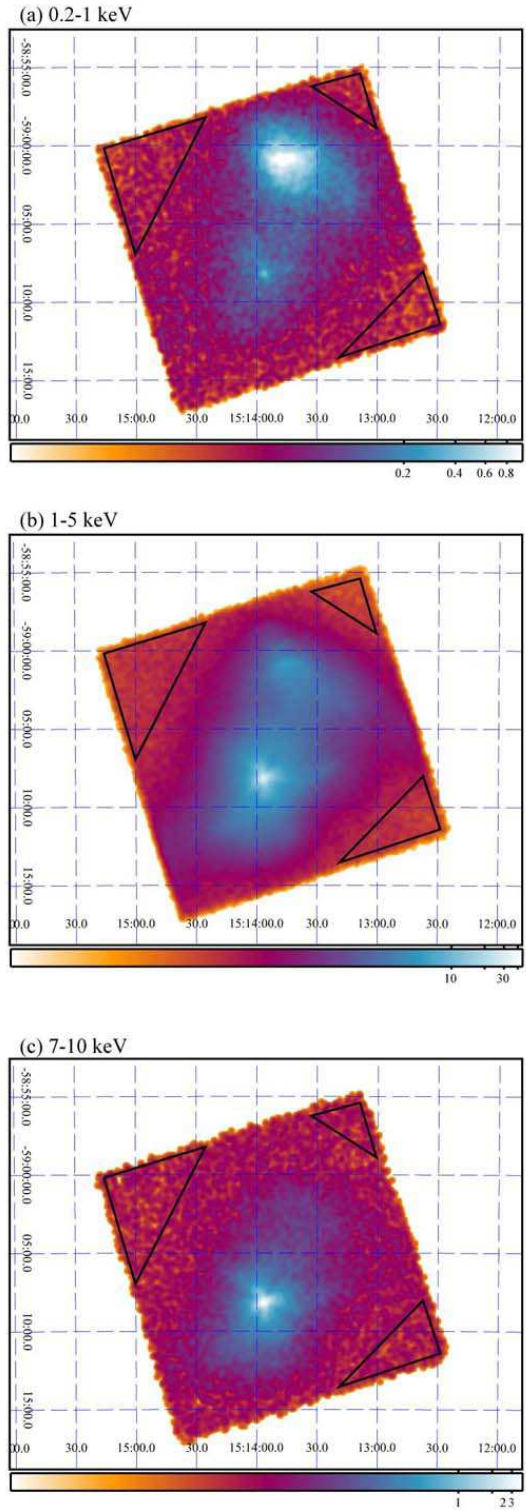}
  \end{center}
  \caption{XIS images in (a) 0.2--1 keV (BI), (b) 1--5 keV
    (FI), and (c) 7--10 keV (FI). Coordinates
    are J2000. For clarity, images are smoothed by a Gaussian
    profile of $\sigma=$15 pixels corresponding to 15''.
    The units on the color scales are count per pixel.
    Solid black lines mark regions utilized in the light curve
    and spectral analyses.
  }\label{fig:img}
\end{figure}

We created XIS images in three representative energy bands from BI 
(back-illuminated, XIS1) and FI (front-illuminated, XIS0, 2, and 3)
cameras as shown in figure \ref{fig:img}.
Two diffuse X-ray emission at the positions of the pulsar and PWN, and 
RCW 89 are observed. The pulsar and PWN emissions located in the south 
east becomes dominant in the hard X-ray band, while the north nebula 
RCW 89 is brighter in the soft band. 
Past higher resolution Chandra images showed similar tendencies (figure 6 
in \cite{gae02}). From these past observations, the pulsar and PWN are 
known to have a hard power law spectrum, while the spectrum of RCW 89 
is dominated by a thermal plasma emission.
To minimize contamination from these X-ray sources, we chose three 
corner regions for the following light curve and spectral analyses. 
Hereafter we call a sum of the three regions a terrestrial diffuse
X-ray emission (TDX) region. A total area of the TDX region is 
41.9 arcmin$^2$.

\section{Light Curve}
\label{sec:lc}

\begin{figure}[p]
  \begin{center}
    \FigureFile(\textwidth,){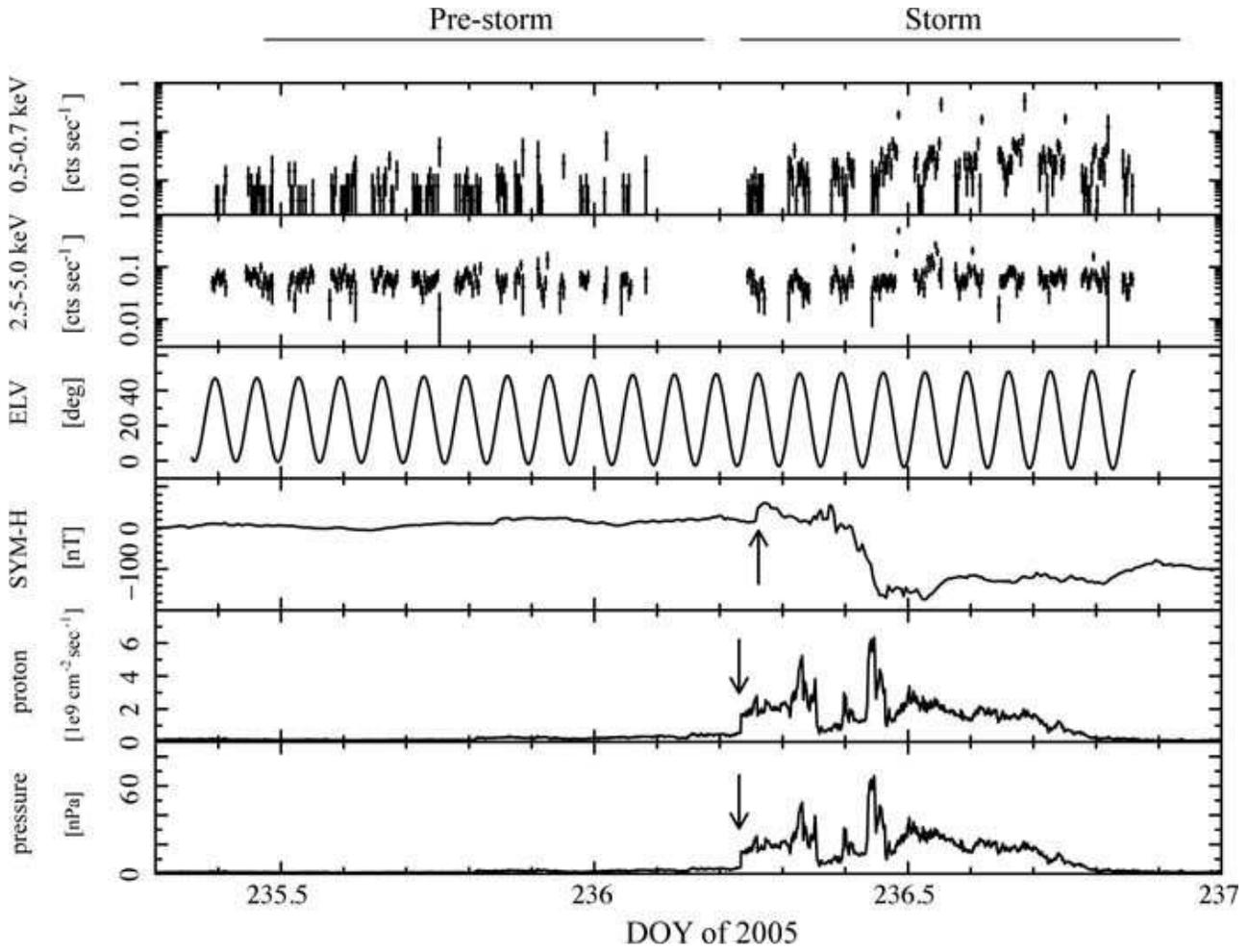}
  \end{center}
  \caption{
 XIS light curves in 0.5--0.7 keV (BI) and 2.5--5 keV (FI), elevation angle from Earth rim, 
 SYM-H index, solar wind proton flux, and solar wind dynamic pressure 
 as a function of DOY in 2005. 
 The vertical errors are 1 $\sigma$ significance.
 The proton flux and dynamic pressure were calculated from WIND SWE data (100 s bin). 
 An arrow in the fourth panel shows a signature of the SSC, while
 those in the fifth and sixths panels indicate increases of the solar wind
 proton flux and dynamic pressure (see text).
  }\label{fig:cur1}
\end{figure}

In figure \ref{fig:cur1}, we plot X-ray light curves extracted
from the TDX region in two energy bands that could potentially 
contain oxygen emission lines from scattering of solar X-rays and/or 
SWCX (0.5--0.7 keV) and for comparison a non-SWCX continuum 
at high energies (2.5--5 keV), similar to \citet{car08}. 
The 0.5--0.7 keV count rate shows a clear enhancement from DOY 
of 236.3 to 236.8. The average rate in the first half of the 
observation (pre-storm period in figure \ref{fig:cur1}) and the 
second half covering (storm period) are 0.0057$\pm$0.0008 
and 0.013$\pm$0.001 cts sec$^{-1}$, respectively. Errors are 
2$\sigma$ significance. 
In contrast, the 2.5--5 keV count rate shows less variability
with the average rates in the pre-storm and storm periods of
0.054$\pm$0.002 and 0.055$\pm$0.002 cts sec$^{-1}$, although
we see some rapid rises before data gaps (e.g., at DOY
of 236.4) and a gradual change around DOY of 236.55.

Several data points both in 0.5--0.7 and 2.5--5 keV show the very
rapid short term increases in the storm period. The scattering 
of solar X-rays is known to vary depending on the elevation angle 
(ELV) of the line of sight from Earth rim. 
Because Suzaku orbits around Earth once per $\sim$90 min, a part
of the observation suffers from Earth occultation. 
The periodic data gaps seen in figure \ref{fig:cur1} correspond to such
occultations. The high count rate bins are seen just before the 
occultation when the ELV angle is relatively small $\sim$5 deg.
Thus, the observed short-term enhancements are most probably 
scattering of solar X-rays. 
A detailed spectral analysis will be done in \S \ref{sec:sec:scattering}.

Using the same time axis, we plot the SYM-H index 
provided by the World Data Center for Geomagnetism, Kyoto, Japan\footnote{http://wdc.kugi.kyoto-u.ac.jp/aeasy/index.html},
and the solar proton flux as measured by WIND\footnote{http://web.mit.edu/space/www/wind\_data.html}.
The SYM-H index is an indicator of the total energy content of 
the ring current almost similar to the hourly Dst index but with 
a high time resolution (1 min) \citep{wan06}.
The SYM-H index exhibits three prototypical periods of the geomagnetic 
storm. The first period is a storm sudden commencement (SSC)
at DOY of 236.26, corresponding to a sharp rise of the SYM-H index, 
when the CME-induced shock arrived at Earth.
The second one is the storm main phase consisting of a dramatic decrease 
of the SYM-H index to a minimum value (-174 nT) when a ring current 
around Earth is built up due to high energy particle enhancements
(e.g., \cite{gon94,miy05}).
The final one is the storm recovery phase seen from DOY of 236.45 
when the SYM-H index had a minimum value.

The solar wind proton data taken with WIND provides 
more information on the solar wind propagation toward Earth. The WIND satellite 
at this observation time was around the sunward L1 Lagrangian point, 
approximately 230 earth radii ($R_{\rm E}$) from Earth. 
From the solar wind proton flux 
and solar wind dynamic pressure 
in figure \ref{fig:cur1}, 
discontinuous rises can be seen at DOY of 236.23 that should be related 
to the CME-induced shock. 
Assuming the average proton speed during the observation of 530 km s$^{-1}$, 
an expected delay of the solar wind to move from the L1 point to Earth 
becomes 0.03 days.
Therefore, an estimated arrival time of the shock at Earth considering 
the delay becomes DOY of 236.26 that coincides well with the SSC, 
i.e., the onset of the geomagnetic storm, seen in the 
ground-based SYM-H index. 

Compared to the SYM-H index and solar wind, the enhancements
of the X-ray light curves in 0.5--0.7 and 2.5--5.0 keV occurred 
after the SSC and the increase of the solar wind proton flux as
well as the dynamic pressure.
Therefore, the X-ray enhancements should be closely related 
to the increased solar wind near Earth. More specifically, 
the soft X-ray enhancement is indicative of the exospheric SWCX 
emission induced by the CME, while that in the hard X-ray
band can be related to increased particle backgrounds.

\begin{figure}[p]
  \begin{center}
    \FigureFile(0.5\textwidth,){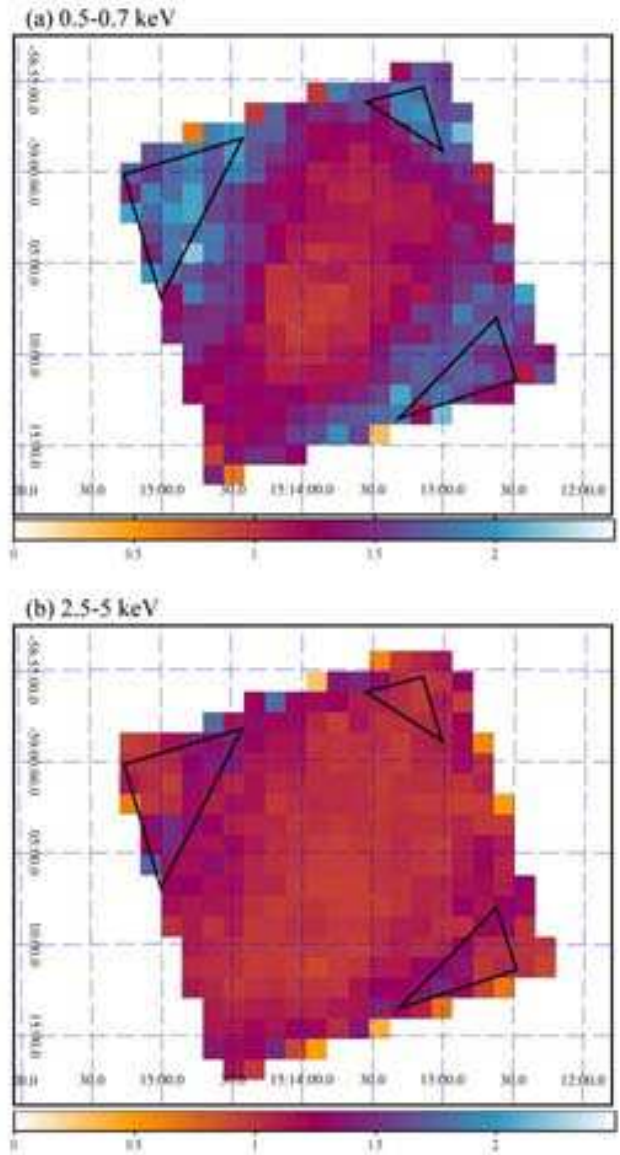}
  \end{center}
  \caption{Ratio maps in (a) 0.5--0.7 keV (BI) and (b) 2.5--5 keV.
    Each image in the storm period is divided by that in the
    pre-storm phase. The exposure time difference is corrected. 
    For clarity, images are binned by 64 pixels.
    The units on the color scales are a ratio.
    Solid black lines mark the TDX region.
  }\label{fig:divimg}
\end{figure}

To check whether this time variability is due to leaked photons from the bright 
X-ray sources, we created ratio maps between the storm and pre-storm 
periods as shown in figure \ref{fig:divimg}. In 0.5--0.7 keV, the TDX 
region shows a factor of $\sim$2 increase, while the bright X-ray sources 
are rather steady. In 2--5 keV, the entire field of view becomes steady.
Therefore, the soft X-ray enhancements must be related to the geomagnetic 
storm and not due to changes in the leaked photons from the bright X-ray sources.

\section{Spectrum}
\label{sec:spec}

\begin{figure}[p]
  \begin{center}
    \FigureFile(0.725\textwidth,){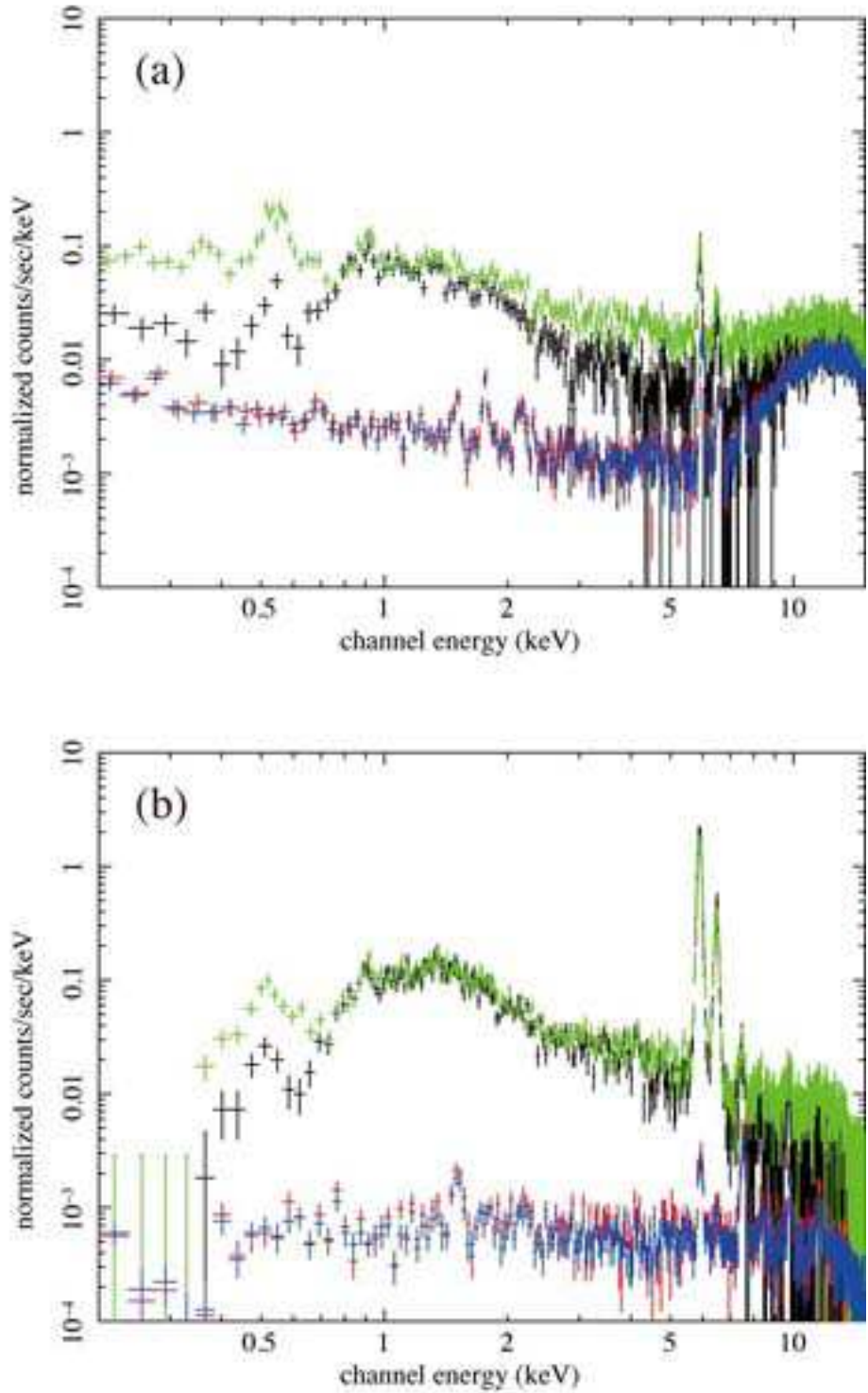}
  \end{center}
  \caption{
 (a) XIS BI and (b) FI spectra of the TDX region before subtracting backgrounds. 
 Black and green data indicate the spectra during the pre-storm and storm periods. 
 Red and blue data are the NXB estimated for the pre-storm and storm periods. 
 Two distinctive emission lines at 5.9 and 6.4 keV are emitted from a calibration $^{55}$Fe sources.
  }\label{fig:spec1}
\end{figure}

We extracted the XIS spectra of the TDX region during the pre-storm 
and storm periods as shown in figure \ref{fig:spec1}. Backgrounds
are not subtracted.
The XIS BI spectrum during the storm period (green) clearly shows 
an excess below $\sim$1 keV compared to that in the pre-storm period 
(black). Signatures of the carbon ($\sim0.3$ keV), nitrogen 
($\sim0.4$ keV) and oxygen (0.5--0.6 keV) emission lines are seen. 
There is also a spectrally smooth excess above $\sim$2 keV.
On the other hand, the FI spectrum during the storm period 
exhibits a clear excess only below 1 keV.
Because the FI CCDs have a higher sensitivity above 1 keV than the BI, 
the hard X-ray excess seen in the BI spectrum will not arise 
from X-ray photons but particle backgrounds.
Ionization particles such as soft protons can produce this spectrally 
smooth, flat, wide-band, whole field-of-view signal (e.g., \cite{car08}). 
These particles are funnelled by the telescope onto the CCD, and 
generate signals. 
The BI CCD is considered to be more susceptible to the particle 
background than the FI because of absence of circuital structures 
upon a pixel.
The increased proton flux seen in the storm period (figure \ref{fig:cur1})
supports this hypothesis. 

To evaluate the instrumental non X-ray background (NXB) in the two 
periods, we utilized the {\tt xisnxbgen} software developed by \citet{taw08}. 
They constructed an NXB database by collecting XIS data when the 
dark Earth covers the XIS field of view.
The NXB spectra (red and blue) of the BI coincide with the TDX spectrum
in the pre-storm period (black) above $\sim7$ keV, where the NXB dominates 
the signal. This is a proof of a reliable estimation of the NXB.
We supposed that the excess of the TDX spectra against the NXB
except for the particle continuum originate from the sky 
background including diffuse galactic and extragalactic emission and 
leaked photons from the bright X-ray sources within the field of view.

\section{Pre-Storm Period}
\label{sec:pre-storm}

We wished to investigate the pre-storm period, in order to quantify the
sky background and contribution from the bright X-ray sources. 
For the spectral fitting, we subtracted the NXB from the TDX spectra,
and created rmf and arf files. We used the {\tt xisrmfgen} and 
{\tt xissimarfgen} programs in the HEAsoft package. For the arf
files, we assumed 
uniform emission from a region of radius 20 arcmin
and corrected obtained X-ray fluxes in spectral fittings for the 
area of the TDX region.
Since we used the $\chi^2$ statistic for our spectral fitting, 
we binned the spectra to a minimum of 20 counts per bin.

In the spectral fitting model, we took into account two 
representative diffuse X-ray backgrounds: the local hot bubble (LHB) 
and the cosmic X-ray background (CXB). 
We assumed the Raymond-Smith thin-thermal plasma model with 
$kT\sim$0.1 keV for LHB based on \citet{sno98}. We also used 
the model Id1 in table 2 of \citet{miy98} for CXB.
In addition, to account for a potential contribution from
the bright X-ray sources within the field of view, we used
models in the works of \citet{tam96} and \citet{gae02}.
The composite spectrum from the pulsar and PWN can be modelled
by an absorbed power law with $N_{\rm H}\sim9.5\times10^{21}$
cm$^{-2}$ and $\Gamma\sim2.0$. The spectrum of RCW 89 is 
represented by an absorbed non-equilibrium ionization (NEI)
model with $N_{\rm H}\sim0.6\times10^{21}$ cm$^{-2}$ and 
$kT\sim0.4$ keV.
From table 1 of \citet{tam96}, we assumed the ionization 
parameter $\tau$ of 6.3$\times10^{10}$ s cm$^{-3}$
and the elemental 
abundances of 0, 0.18, 0.49, 0.26, 1.12, and 1.1$\times10^{-2}$ 
solar for O, Ne, Mg, Si S, and Fe, respectively. 
The solar elemental abundance table by \citet{and89} was used.
To estimate the contribution from these sources, we let only 
the normalizations of these four models free. 

\begin{figure}[p]
  \begin{center}
    \FigureFile(0.725\textwidth,){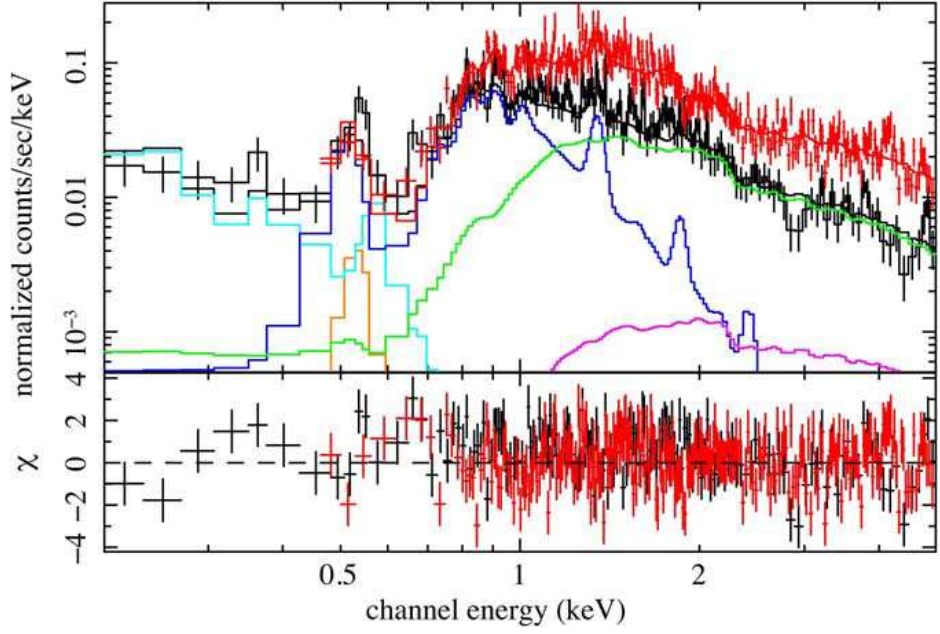}
  \end{center}
  \caption{
 Background-subtracted 
 XIS BI (black) and FI (red) spectra of the TDX region in the pre-storm period.
 In the top panel, the solid line is the best-fit model. For clarity, model 
 components are shown only for the BI spectrum and color coded: power law (green), 
 NEI (blue), LHB (cyan), CXB (magenta), and Gaussian (orange). The best-fit 
 parameters are summarized in table \ref{tbl:spec2}.
  }\label{fig:spec2}
\end{figure}

After fitting tests, 
we found that the surface brightness of the CXB component 
is $\sim$5 times smaller than the observed continuum. 
Hence, we fixed all the spectral parameters of CXB 
as described in \citet{miy98}.
We also found that there is a residual around 0.5--0.6 keV, most 
probably from the neutral oxygen line, and hence added a Gaussian model. 
Because the spectral fitting was not acceptable at this stage, we 
finally let the temperature and abundance of Fe in the NEI model free. 
The obtained result is shown in figure \ref{fig:spec2} and table 
\ref{tbl:spec2}. Below the errors are 90\% confidence range unless
otherwise are noted. The model represents the data well with a
$\chi^2$/d.o.f. of 1.41.
It is evident from figure \ref{fig:spec2} that the soft sky 
background model dominates the spectrum below 0.7 keV, while
the bright X-ray source models reproduce the data above 0.7 keV.
The obtained surface brightness of the LHB, 
1.7$\times$10$^{-15}$ 
erg s$^{-1}$ cm$^{-2}$ arcmin$^{-2}$, is consistent with the 
previous observations \citep{sno98,ham07,ezo09}.
The best-fit temperature of the NEI model also coincides with the
temperature of RCW 89 (0.39$\pm$0.05 keV in \cite{tam96}) within
errors, while the iron abundance is an order of magnitude higher.  

We estimated that a contribution from the PWN and RCW 89 to the TDX 
region is $\sim$2 and 7 \% of their total emission, respectively.
Here we cited \citet{gae02} and \citet{tam96} for the total X-ray 
fluxes from the PWN and RCW 89, respectively. 
These numbers are reasonable if we consider the angular response of 
the Suzaku X-ray telescope (see figure 12 in \cite{ser07}). When a 
point source is located at the center of XIS, 2$\sim$3\% photons 
fall outside $r=6$ arcmin. Because the PWN and RCW89 are extended, 
more photons will fall within the TDX region, which is located 
$6\sim10$ arcmin from the aimpoint. 
The larger percentage of the thermal NEI component compared to the 
power-law can be explained by an extra sky background, a so-called 
transabsorption emission (TAE) \citep{kun08,yos09}.
The TAE is considered to arise from distant part of the galaxy, above 
or beyond the bulk absorption of the galactic disk. Its spectrum is 
described by a thermal emission model with $kT\sim$ 0.3 keV, comparable 
to the temperature of RCW 89. Its flux varies depending on the line of 
direction \citep{yos09}. The TAE may also explain the Fe line abundance 
of the NEI component.

\section{Storm Period}
\label{sec:storm}

\subsection{Scattering of solar X-rays}
\label{sec:sec:scattering}

\begin{figure}[p]
  \begin{center}
    \FigureFile(0.725\textwidth,){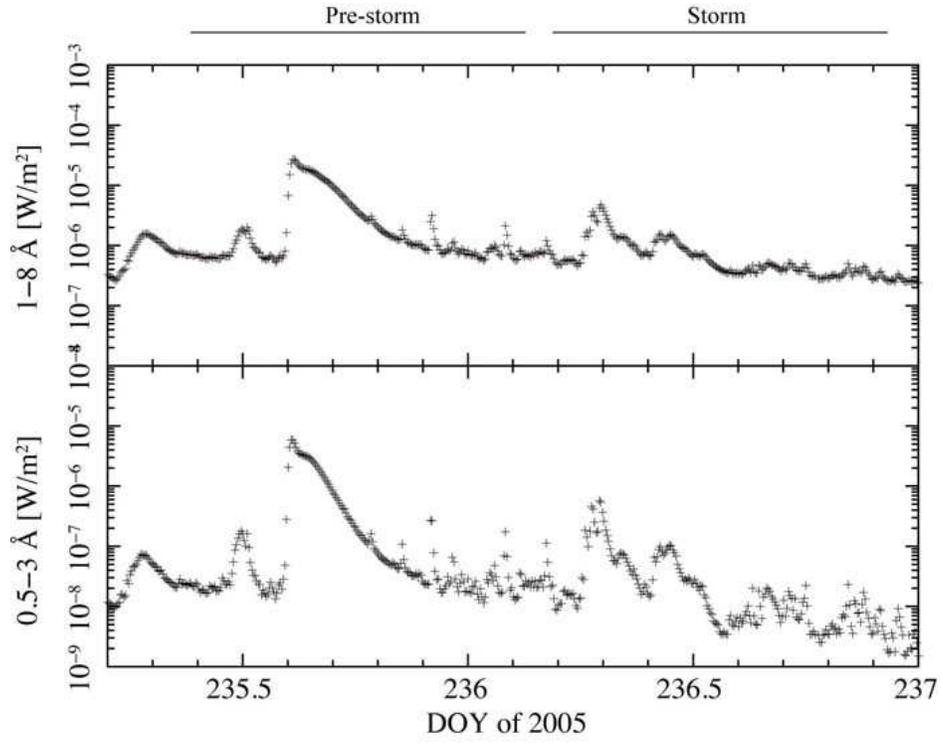}
  \end{center}
  \caption{
 Solar X-ray fluxes in 1--8 \AA~ and 0.5--3 \AA~ taken with GOES. 
Two top bars indicate the time range of the pre-storm and storm periods.
  }\label{fig:goes}
\end{figure}

Having known that the pre-storm spectrum is dominated by the soft sky
background below 1 keV and by the contamination from the bright 
X-ray sources above 1 keV, we proceeded to study the TDX emission 
in the storm period. 
As suggested from figure \ref{fig:spec1}, the major difference between 
the storm and pre-storm period concentrates below 1 keV. Considering 
the higher sensitivity of the BI than that of the FI, below we only analyze 
the XIS BI spectrum. 
Firstly we investigated the scattering of solar X-rays 
that was suggested by the X-ray light curve (figure \ref{fig:cur1}) 
and the spectrum during the pre-storm period (figure \ref{fig:spec2}).
In figure \ref{fig:goes}, the GOES12 solar X-ray fluxes in two
energy bands are plotted. The data were taken from the 
National Geophysical Data Center\footnote{http://goes.ngdc.noaa.gov/data/avg/2005/}.
An M-class flare occurred at DOY of 235.6 during the pre-storm period. 
The average X-ray fluxes during the pre-storm period in 1--8 \AA~ or 
1.55--12.4 keV and 0.5--3 \AA~ or 4.13--24.8 keV 
were 3.4$\times10^{-6}$ and 4.2$\times10^{-7}$ W m$^{-2}$,
while those during the storm periods were
8.0$\times10^{-7}$ and 4.2$\times10^{-8}$ W m$^{-2}$, respectively.
Although the level of the solar X-rays are lower in the storm period, 
atmospheric expansion associated with the CME can cause an increase 
of the fluorescence line(s) in the storm period.

\begin{figure}[p]
  \begin{center}
    \FigureFile(0.725\textwidth,){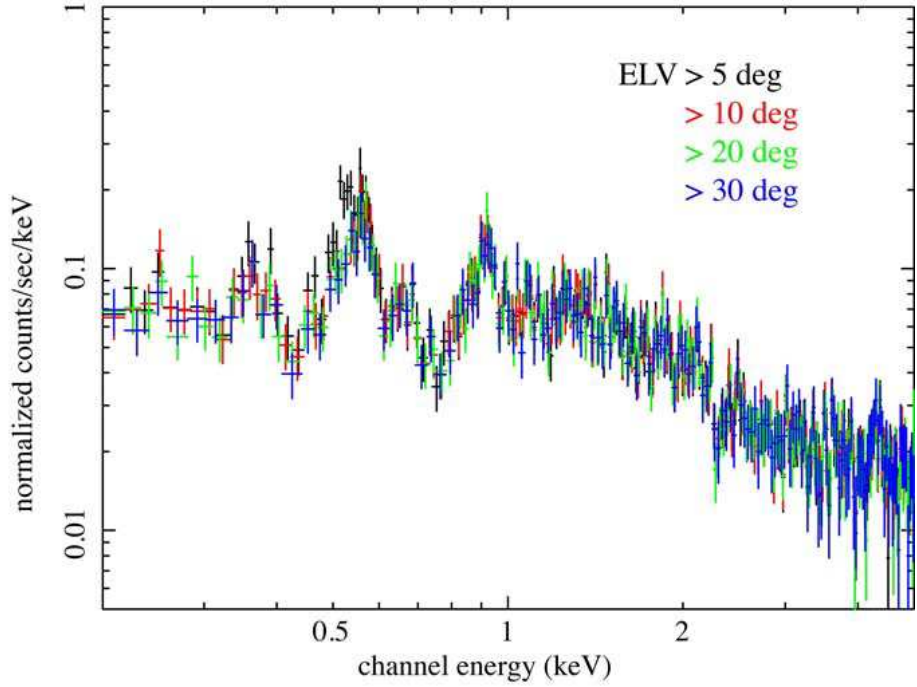}
  \end{center}
  \caption{
 Background-subtracted BI spectra during the storm period. 
 Different colors correspond to different ELV criteria.
  }\label{fig:spec3}
\end{figure}

\begin{figure}[p]
  \begin{center}
    \FigureFile(0.725\textwidth,){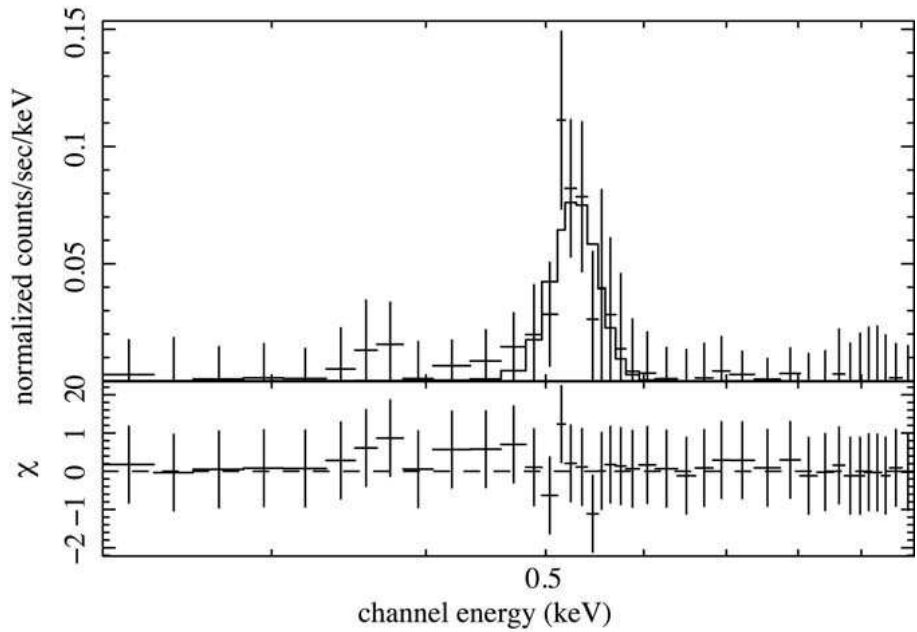}
  \end{center}
  \caption{
 XIS BI spectrum in the storm period when the ELV angle is 5$\sim$10 deg.
 The solid line in the upper panel is the best-fit Gaussian model. 
 The lower panel shows the data residuals from the model. 
 The parameters are summarized in table \ref{tbl:spec4}.
  }\label{fig:spec4}
\end{figure}

With this thought in mind, we changed the data filtering criteria on 
the ELV angle. The default criteria for the screened data is ELV$>$5 deg. 
In order to investigate a potential difference, we adopted more stringent 
ELV criteria (ELV$>$10, 20, and 30 deg). 
Figure \ref{fig:spec3} shows the XIS BI spectra for the four ELV criteria. 
An increase of a neutral oxygen line 
at $\sim$0.53 keV
is recognized in the ELV$>5$ deg spectrum,
while there is no significant change between ELV$>10$ and $>30$ deg in 0.2--5 keV.
We obtained the BI spectrum when ELV is 5$\sim$10 deg by subtracting
the ELV$>10$ deg spectrum from that with ELV$>5$ deg. Figure \ref{fig:spec4} 
is the spectrum after this subtraction. The data is modelled with a single 
narrow Gaussian model. The parameters are 
summarized in table \ref{tbl:spec4}. 
The line center energy is consistent with the neutral oxygen K line
of 0.524 keV \citep{tom09}\footnote{http://xdb.lbl.gov/}.

By adopting the ELV criteria of $>10$ deg, we also noticed that the 
shape of the light curve also changes. The high count rate events 
are suppressed. 
Our result warns that, even if the solar X-ray flux is not high, 
the standard Suzaku data screening criteria of ELV $>5$ deg is not 
sufficient to remove the scattering component, when the atmospheric 
density rises after CMEs.

\subsection{Solar wind charge exchange}
\label{sec:sec:swcx}

\begin{figure}[p]
  \begin{center}
    \FigureFile(0.725\textwidth,){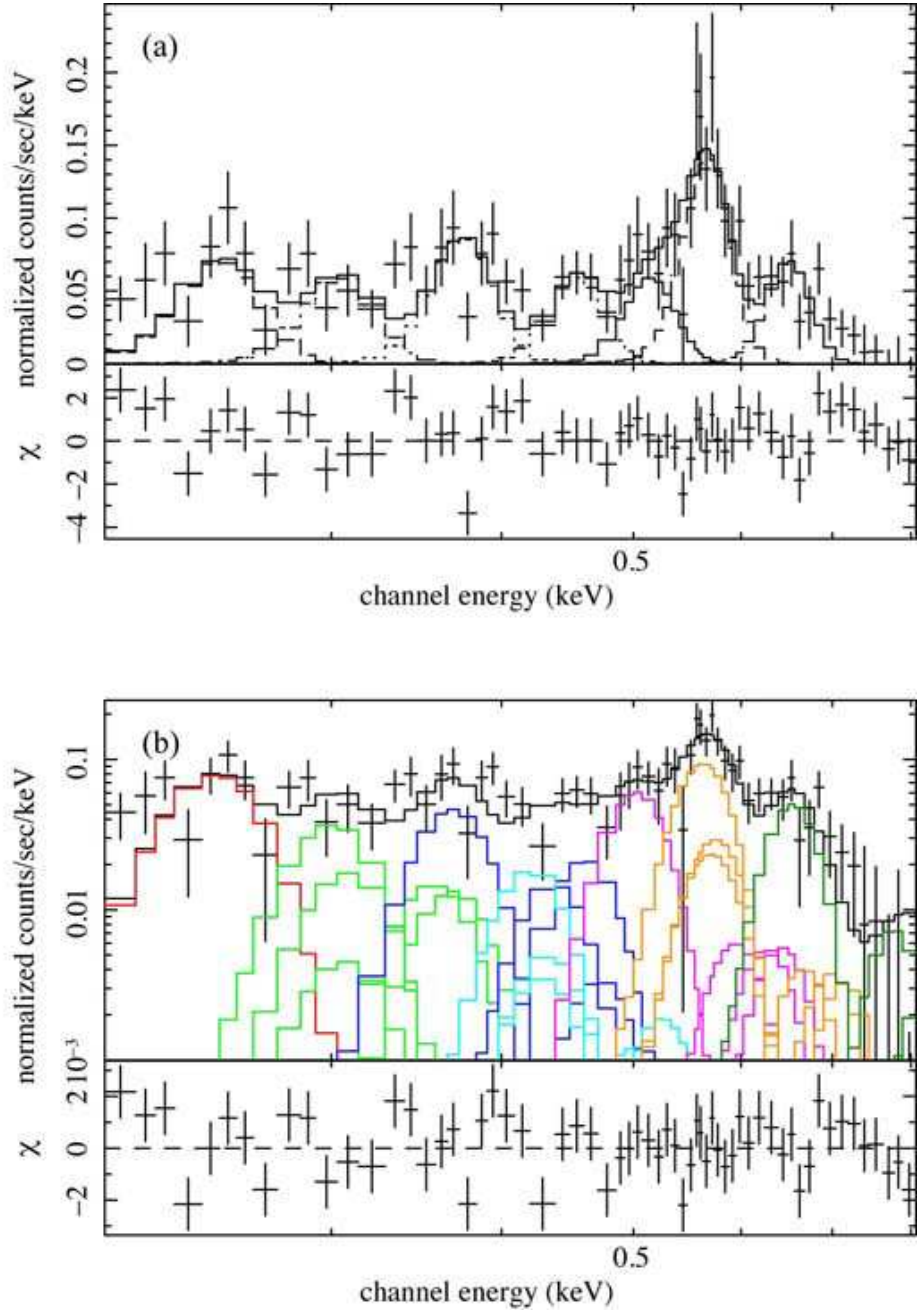}
  \end{center}
  \caption{
 XIS BI spectrum in the storm period when the ELV angle is $>10$ deg. 
 The pre-storm period spectrum is subtracted as a background. 
 In the panel a, the seven Gaussian model is used and the 
 parameters are listed in table \ref{tbl:spec5}.
 In the panel b, the SWCX model is used and the parameters
 are in table \ref{tbl:spec5-2}. The lines due to C, N, and 
 O are color coded: 
 C\emissiontype{V} (green), 
 C\emissiontype{VI} (blue), 
 N\emissiontype{VI} (cyan), 
 N\emissiontype{VII} (magenta), 
 O\emissiontype{VII} (orange), and
 O\emissiontype{VIII} (dark green).
  }\label{fig:spec5}
\end{figure}

We then analyzed possible enhancements by the exospheric SWCX. 
To remove the solar X-ray scattering effect, we screened the data 
with the ELV$>$10 deg criteria. In order to see differences 
between the pre-storm and storm periods, we subtracted the 
pre-storm period spectrum from the storm period. In figure 
\ref{fig:spec5} (a), we plot the obtained spectrum.
It is evident from figure \ref{fig:spec5} (a) that the excess 
clearly contains emission lines between 0.2 and 0.7 keV. The 
spectrum is built up from a series of emission lines from carbon, 
nitrogen, and oxygen lines. 
The spectrum was modelled with seven narrow Gaussians, but
the fitting was not statistically acceptable ($\chi^2$/d.o.f. 
of 1.91). In table \ref{tbl:spec5}, we list the obtained 
parameters. 

We then tested a theoretical SWCX emission line model 
developed by \citet{bod07} (table 8.2), which accounts 
for the relative emission cross sections 
for the lines from each of  several ions, specifically
C\emissiontype{V}, C\emissiontype{VI}, N\emissiontype{VI}, 
N\emissiontype{VII}, O\emissiontype{VII}, and O\emissiontype{VII}.
In total, 33 lines are involved in this model. 
\citet{car10} have successfully fitted the exospheric SWCX spectrum 
taken with XMM-Newton.
We used the tabulated values for a velocity of 600 km s$^{-1}$,
because the average solar wind velocity during the storm period
is 530 km s$^{-1}$.
We allowed the six normalizations of the ions to be free.  
We also added an extra Gaussian to reproduce the lowest 
energy emission line around 0.25 keV, which is not included
in the Bodewits's SWCX model.

In figure \ref{fig:spec5} (b), we plot the fit results. 
This SWCX model reproduced the data significantly better 
than the simple sum of Gaussians with $\chi^2$/d.o.f. of 1.46. 
The obtained parameters are listed in table \ref{tbl:spec5-2}.
This acceptable fitting supports the idea that the soft X-ray enhancements
during the storm period is due to the exospheric SWCX.
The emission line identified around 0.45 keV mainly arises 
from the $n=$4 to 1 transition of C VI at 459 eV, which is 
hardly seen in a normal astrophysical plasma and hence is another 
line of evidence for the exospheric SWCX \citep{fuj07}.

\begin{figure}[p]
  \begin{center}
    \FigureFile(0.725\textwidth,){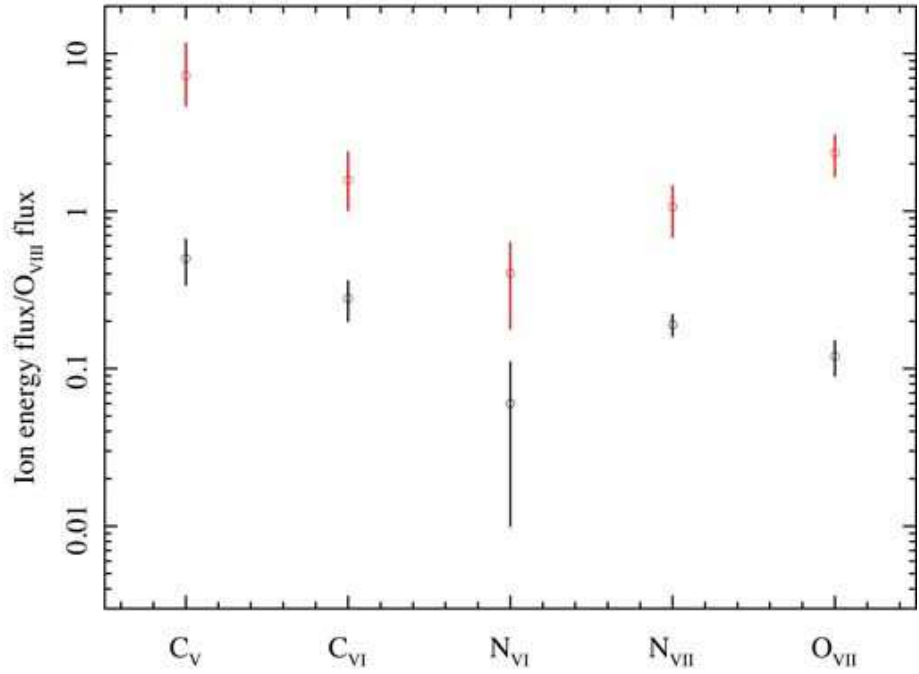}
  \end{center}
  \caption{Line energy flux ratio to O\emissiontype{VIII}. Red and
  black data points indicate the best fit parameters of this observation
  and \citet{car10}, respectively.
  }\label{fig:ratio}
\end{figure}

Figure \ref{fig:ratio} shows the best-fitting SWCX line energy flux 
ratio to O\emissiontype{VIII} compared to the XMM-Newton
observation of CME-induced exospheric SWCX by \citet{car10}.
While the relative flux ratios from C\emissiontype{V} to 
N\emissiontype{VII} are similar, the absolute values in this
Suzaku observation are an order of magnitude higher.
Indeed, the O\emissiontype{VIII} line was prominent in 
\citet{car10}, while O\emissiontype{VII} dominates in this observation. 
We will examine whether this difference can be explained by
differences in the incident solar wind O$^{7+}$ and O$^{8+}$ 
fluxes and SWCX cross sections in \S \ref{sec:int}.

To investigate a potential influence of the hard continuum 
originating from the particle background, we fitted the 1--5 keV 
BI spectrum in the storm period after the subtraction 
of the pre-storm spectrum with a power-law. 
The fitting was acceptable with $\chi^2$/d.o.f. of 229/190$=1.21$. 
The best-fit photon index and normalization were 
$0.08^{+0.17}_{-0.19}$ and $26^{+6}_{-5}$ photon s$^{-1}$ cm$^{-2}$ str$^{-1}$ 
(line unit, LU) at 1 keV, respectively.
We then extrapolated this power law continuum into the 
low energy band. We again fitted the BI spectrum shown
in figure \ref{fig:spec5} with a sum of the power law
and the SWCX line models.
We fixed the photon index and normalization of the 
power-law component. We then found that all the 
line normalizations are reduced by 5$\sim20$ \% but 
almost all of these parameters are within 90\% confidence 
range listed in table \ref{tbl:spec5-2}. 
Therefore, 
the particle continuum may
be minor in the low energy band, in comparison with the SWCX lines.

\section{Time Correlation}
\label{sec:cor}

\subsection{O\emissiontype{VII} line vs O$^{7+}$ ion}
\label{sec:sec:ovii1}

\begin{figure}[p]
  \begin{center}
    \FigureFile(\textwidth,){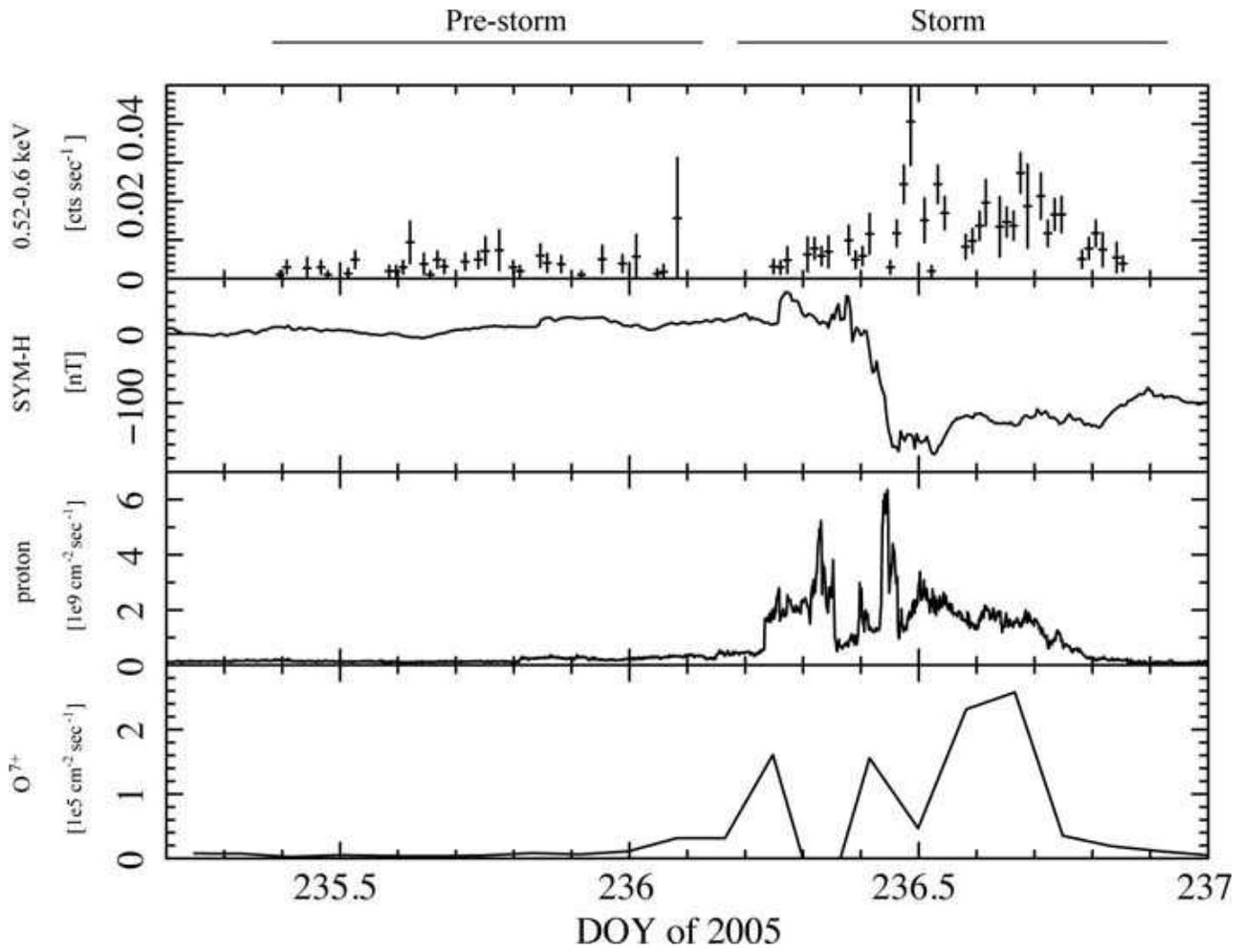}
  \end{center}
  \caption{XIS BI 0.52--0.6 keV light curve filtered with ELV$>$10 deg, 
  the SYM-H index, the solar wind proton flux, and the solar wind O$^{7+}$ 
  flux from level 2 ACE SWICS data (2 hr bin).
  }\label{fig:cur2}
\end{figure}

An X-ray flux of the exospheric SWCX emission can be estimated
from an incident solar wind flux and a neutral column density 
of the exosphere in the line of sight.
To assist in this analysis, we utilized the solar wind oxygen
ion flux as measured by ACE SWICS\footnote{http://www.srl.caltech.edu/ACE/ASC/level2/index.html}. 
We compared the O$^{7+}$ flux with the XIS BI 0.52--0.6 keV O\emissiontype{VII} 
count rate.
To eliminate the contribution of the neutral oxygen line, 
the XIS data are filtered with the criteria of ELV$>$10 deg.
Figure \ref{fig:cur2} shows the XIS BI 0.52--0.6 keV light curve
compared to the SYM-H index, WIND proton flux, and ACE O$^{7+}$
ion flux. As we wrote in \S \ref{sec:sec:scattering}, there is 
no sign of very high count rate bins due to the scattering of 
solar X-rays after we filtered the data with the ELV$>10$ deg
criterion.
Similar to figure \ref{fig:cur1}, the X-ray enhancement
started after the onset of the geomagnetic storm at DOY of 
236.25, supporting that the X-ray rise is due to the increased
solar wind near Earth and the resultant exospheric SWCX.

Before checking the relation between the X-ray and ion fluxes,
we conducted a cross-correlation analysis to know of any time
delay between these two data.
This procedure requires that both light curves are taken in 
equally-spaced time intervals. Unfortunately, the ACE ion data 
has rather low time resolution
 (2 hr average) and hence we had to 
bin the XIS BI curve into 8192 s bin. To consider different
time bins of these two curves, we interpolated the ACE data 
to match the XIS curve in the same way as in \citet{ezo10}. 
We then utilized the {\tt crosscorr} software in the HEAsoft
package with the default mathematical algorithm and 
normalization method. 

\begin{figure}[p]
  \begin{center}
    \FigureFile(0.725\textwidth,){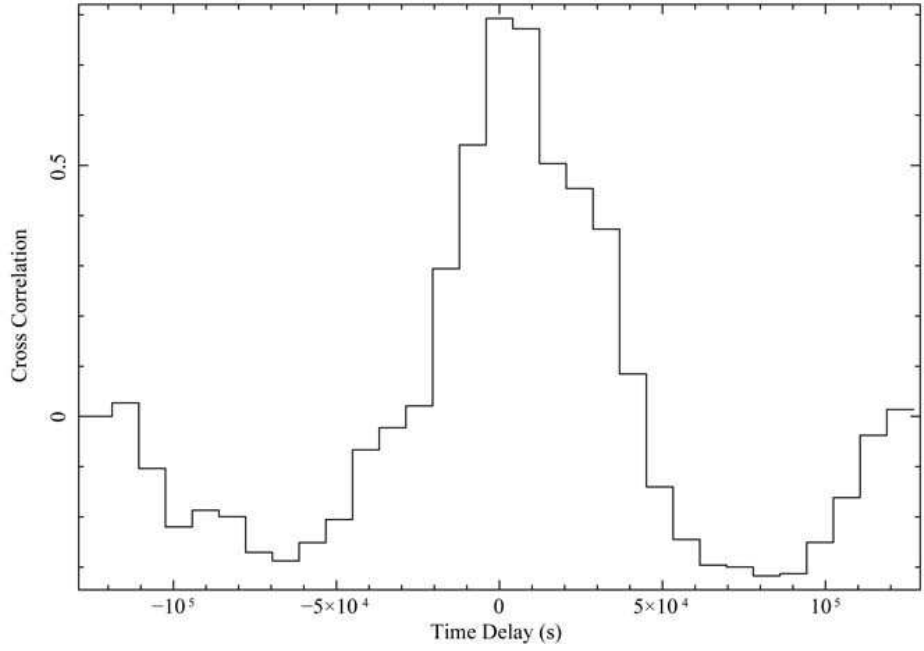}
  \end{center}
  \caption{
 Cross correlation between the XIS BI O\emissiontype{VII} count rate
 and ACE O$^{7+}$ flux. A positive delay means
 that the ACE data leads the XIS BI.
  }\label{fig:cross1}
\end{figure}

In figure \ref{fig:cross1}, we plot the cross correlation. 
The correlation coefficient has a peak (0.79) at a time delay of 
0$\sim$8192 s. A null hypothesis probability is $1\times$10$^{-3}$, 
corresponding to $\sim3\sigma$ significance. 
The obtained time delay coincides with the expected value 
of $\sim1$ hr. 
Because the ACE satellite orbits at the L1 point and Suzaku 
is in the low Earth orbit, about 1 hr time delay is expected 
between these two data (see \S \ref{sec:lc}).

\begin{figure}[p]
  \begin{center}
    \FigureFile(0.725\textwidth,){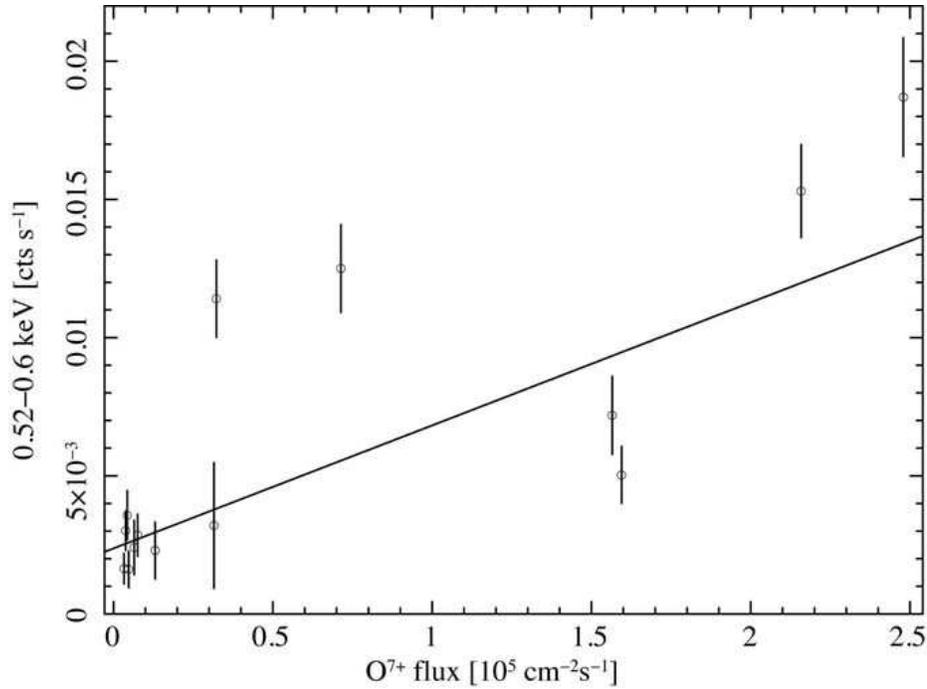}
  \end{center}
  \caption{
 Correlation between the XIS BI O\emissiontype{VII} count rate
 and ACE O$^{7+}$ ion flux. The vertical error bar
 is $1\sigma$ significance. The solid curve is the 
 best-fit linear function. 
  }\label{fig:corr1}
\end{figure}

In figure \ref{fig:corr1}, we plot a relation between the 
O\emissiontype{VII} count rate and the solar wind O$^{7+}$ ion flux. Because the estimated
time delay is consistent with 1 hr, i.e., within the bin size of the light 
curve, we did not correct the data for the time delay.
We can see that two quantities are correlated.
We fitted the data with a linear function as shown in a solid line,
although the fitting is not acceptable ($\chi^2$/d.o.f. of 86.9/14).
We checked the relation assuming the 
time delay of 8192 s, and found a similar increase.

\subsection{O\emissiontype{VII} line vs proton}
\label{sec:sec:ovii2}

\begin{figure}[p]
  \begin{center}
    \FigureFile(0.725\textwidth,){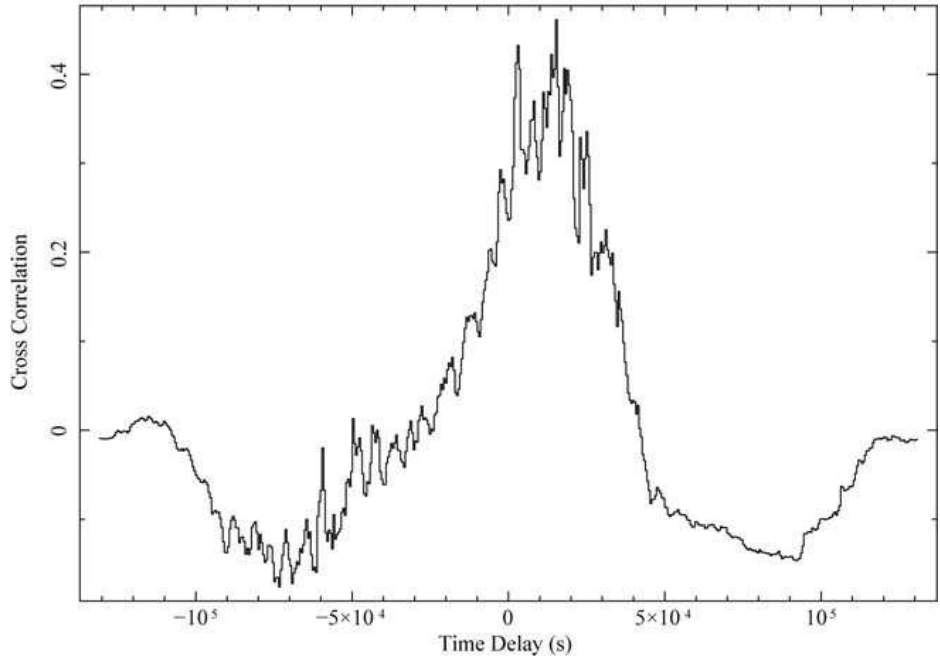}
  \end{center}
  \caption{
 The same as figure \ref{fig:cross1} but for the XIS BI O\emissiontype{VII} 
 count rate and WIND proton flux.
  }\label{fig:cross2}
\end{figure}

The sparse O$^{7+}$ data motivated us to utilize the high time resolution 
proton data. Using the proton data (100 s average), we checked the
time delay and then checked the correlation between the solar wind proton 
flux and O\emissiontype{VII} count rate, in the same way as the ion data.
Figure \ref{fig:cross2} shows the cross correlation. A broad peak 
is seen in the range of 0$\sim3\times10^4$ sec. This range covers the 
expected time delay of 1 hr, although the peak is slightly shifted at 
the positive side ($\sim1.5\times10^4$ sec), suggesting that there 
could be an additional positive delay. 

\begin{figure}[p]
  \begin{center}
    \FigureFile(0.725\textwidth,){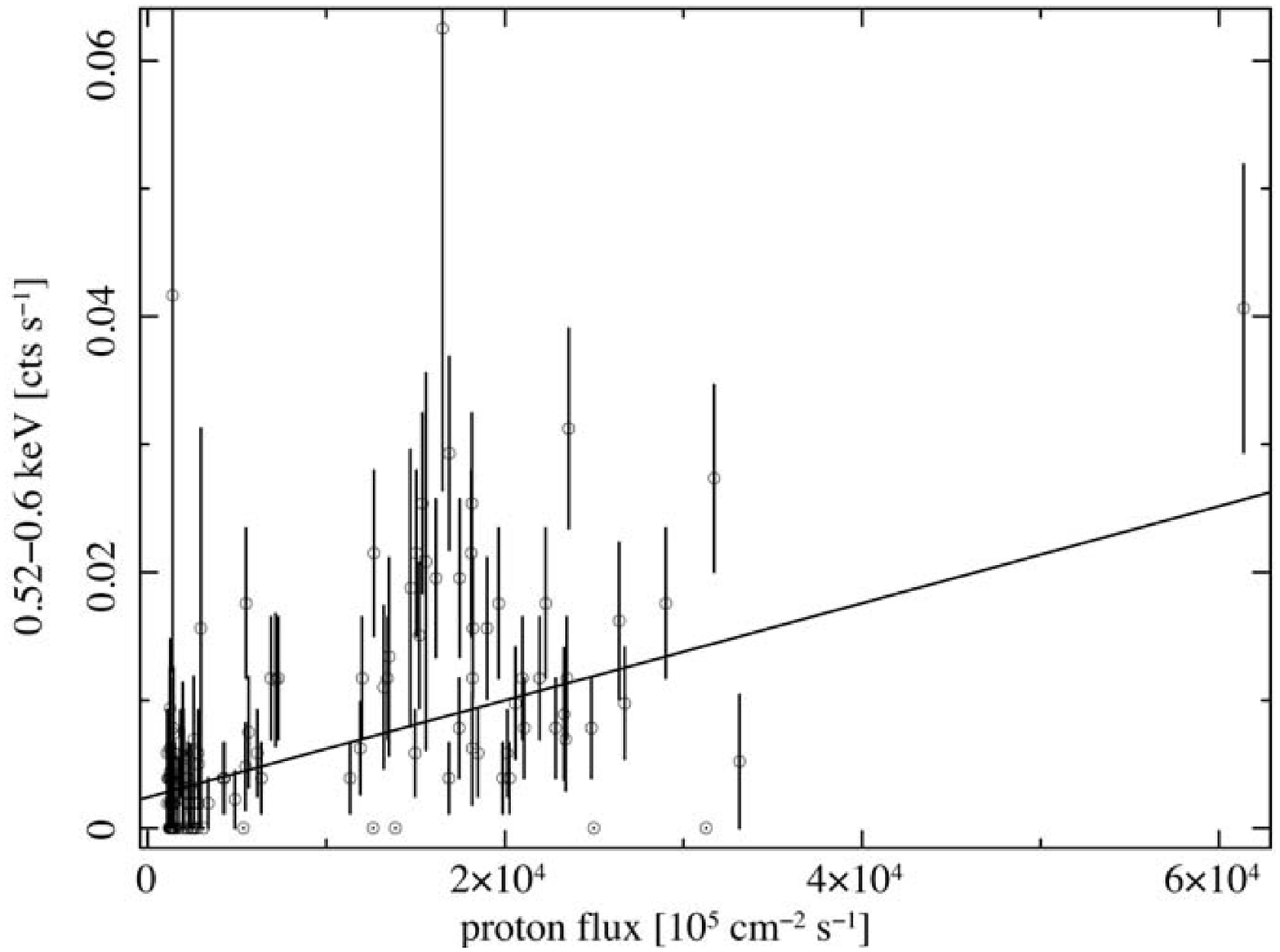}
  \end{center}
  \caption{
 Correlation between the XIS BI O\emissiontype{VII} count rate
 and WIND proton flux. The vertical error bar
 is $1\sigma$ significance. The solid curves are the 
 best-fit linear function. 
  }\label{fig:corr2}
\end{figure}

Figure \ref{fig:corr2} displays a correlation between the O\emissiontype{VII} 
count rate and the solar wind proton flux. A time delay of 1 hr is assumed based on the satellite
positions. The data is represented by a linear function expressed as,
\begin{equation}
C_{\rm XIS}~ {\rm [cts~ s^{-1}]} = a \times C_{\rm proton}~ {\rm [10^5~ cm^{-2}~ s^{-1}]} + b, 
\end{equation}
where $a$ represents the SWCX emissivity and $b$ is an offset emission 
due to the instrumental and sky background.
The fitting was acceptable with $\chi^2$/d.o.f. $=$119.5/106. 
The best fit $a$ and $b$ are $3.8\pm0.7\times10^{-7}$ and
$2.4\pm0.7\times10^{-3}$, respectively.
The positive $b$ suggests that, even if the solar wind flux is zero, 
there remains an 
certain level of background emission. Since the NXB is minor as seen in figure 
\ref{fig:spec1}, a large part of $b$ must arise from the sky background.
We converted $b$ into the photon flux in units of LU based on the 
O\emissiontype{VII} fitting in figure \ref{fig:spec5}. 
Then, $b$ corresponds to $10\pm3$ LU. 
\citet{yos09} reported that the O\emissiontype{VII} line intensity from 
the soft X-ray background consisting of LHB, TAE and heliospheric SWCX 
ranges from 2 to 10 LU. Thus, we can interpret $b$ as the sum of these 
sky backgrounds.

\section{Expected Intensity}
\label{sec:int}

\subsection{Scattering of solar X-rays}
\label{sec:sec:scat}

The expected oxygen line intensity of the fluorescent scattering of solar X-rays can 
be estimated from the equation below.
\begin{equation}
P_{\rm scat} = \frac{1}{4\pi}\int_{E_0} \sigma_{\rm scat} (E)~ \eta~ P_{\rm Sun}(E)~ 
N_{\rm O+O2}~ dE~{\rm [LU]},
\end{equation}
where $E_0$ is the minimum X-ray energy needed for the fluorescent scattering, 
$\sigma_{\rm scat}(E)$ is the photoelectric absorption cross section of oxygen 
per an atom as a function of X-ray energy $E$, $\eta$ is the fluorescent yield, 
$P_{\rm Sun}$ is the incident X-ray photon flux, and $N_{\rm O+O2}$ is the column 
density of oxygen atoms and molecules in the line of sight.

We found $\sigma_{\rm scat}(E)$ from \citet{hen93}\footnote{http://henke.lbl.gov/optical\_constants/} 
and used $\eta$ of 0.0083 from \citet{kra79}. We estimated the solar X-ray spectrum
based on the GOES data (figure \ref{fig:goes}) as below. The solar X-ray spectrum 
is known to be variable and can be expressed by a multi-temperature thin thermal 
plasma model with the temperature of 0.1$\sim$3 keV (e.g., \cite{phi99}). 
For simplicity, we assumed a single temperature plasma and estimated the plasma 
temperature from the ratio of the two-band GOES X-ray fluxes. We utilized the 
{\tt MEKAL} model in the {\tt xspec} software package to simulate the solar X-ray spectrum.
The average ratios during the pre-storm and storm periods indicate plasma temperatures 
of 1.35 keV and 0.98 keV, respectively. We tuned the normalization of the 
{\tt MEKAL} model so that the the GOES two-band fluxes are reproduced.

We then multiplied the cross section, fluorescent yield, and solar X-ray spectrum,
and then integrated the term as a function of energy from 1 keV, i.e., above the 
oxygen K-edge (0.54 keV), to 25 keV. 
We here used 1 keV as a first trial taking into account the lowest energy of
the GOES X-ray data (1.55 keV) used to estimate the solar X-ray spectrum.
We divided the observed flux by this integral
and obtained the necessary $N_{\rm O+O2}$.
In the pre-storm period, the observed oxygen line flux was $<2.4$ LU 
(table \ref{tbl:spec2}). 
Then, the necessary $N_{\rm O+O2}$ becomes 
$<3\times10^{15}$ cm$^{-2}$. In the storm period, the line flux was increased 
by $14$ LU (table \ref{tbl:spec4}). 
This corresponds to rise of $N_{\rm O+O2}$ by $4\times10^{16}$ cm$^{-2}$. 
Thus, a factor of $>$10 increase in the column density is suggested. 
This increase is due to the fact that the solar X-ray intensity
was 2 times stronger in the pre-storm period, while the observed neutral 
oxygen line flux was $>$5 times weaker compared to that in the storm period.

For comparison, we estimated $N_{\rm O+O2}$ by integrating the atmospheric 
neutral oxygen atom and molecules of the Sun lit atmosphere in the line of 
sight, using the NRLMSIS2000 empirical model
\citep{pic02}\footnote{http://www.nrl.navy.mil/research/nrl-review/2003/atmospheric-science/picone/}.
This is an empirical temperature and density model of the Earth's atmosphere 
from ground to space. We calculated the geodetic latitude, longitude, and 
altitude of the Suzaku satellite, and integrated the O$+$O$_2$ density as 
a function of time. The density is estimated  by a 10 km step from the 
altitude of Suzaku ($\sim570$ km during the observation) to 1000 km, and 
then integrated by multiplying the density at each place by 10 km. 

The NRLMSIS2000 model provides us with $N_{\rm O+O2}$ as a function of time. 
We extracted the average $N_{\rm O+O2}$ in the pre-storm and storm periods.
Because the scattering component is seen only in low ELV angles (5$\sim$10 deg),
we averaged $N_{\rm O+O2}$ over this ELV range. 
The estimated $N_{\rm O+O2}$ in the pre-storm and storm periods 
were $\sim5\times10^{16}$ and $\sim7\times10^{16}$ cm$^{-2}$, 
respectively. This theoretical column density in the storm period coincides 
with the observation-based value within a factor of 2, while that in the pre-storm period 
is 20 times larger. 

There are two major uncertainties in our estimation. The first one is the assumed lower 
energy of the solar X-ray spectrum (1 keV). If we integrate the spectrum down to 0.54 keV, 
$N_{\rm O+O2}$ decreases to $<$0.5$\times10^{15}$ and 7$\times10^{15}$ cm$^{-2}$ for 
the pre-storm and storm phases, respectively, owe to the increased solar X-ray photons. 
If we adopt the lower energy of the GOES data (1.55 keV), $N_{\rm O+O2}$ become 
$<$3$\times10^{16}$ and 1$\times10^{17}$ cm$^{-2}$. These numbers can be considered as 
conservative upper limits. 
The other is the NRLMSIS2000 model itself. Comparison of the model with the satellite drag 
data indicates that there can be a factor of 2 difference \citep{pic02}. For example, 
from the incoherent scatter in the polar upper atmosphere, \citet{oga09} found that a H 
density at 600 km altitude is roughly 
five times higher than the model. Therefore, a factor of 2$\sim$5 error can reside in 
the model. Thus, the observed discrepancy can be reduced by these uncertainties, although
the increased $N_{\rm O+O2}$ in the geomagnetic storm, which was not reproduced in the 
NRLMSIS2000 model, will be left as a future issue.

\subsection{Solar wind charge exchange}
\label{sec:sec:swcx}

Next, we estimated the intensity of the exospheric SWCX emission. 
This is expressed by the equation below.
\begin{equation}
P_{\rm SWCX} = \frac{1}{4\pi} \alpha~ P_{\rm SW}~ 
N_{\rm H}~ {\rm [LU]}, 
\label{eq:swcx}
\end{equation}
where $\alpha$ accounts for the charge exchange cross section and 
line emission probability, $P_{\rm SW}$ is the incident solar wind 
flux, and $N_{\rm H}$ is the column density of target hydrogen atoms in 
the Earth's exosphere.

To calculate $N_{\rm H}$ for the observed O\emissiontype{VII} line in
the storm period, we used $\alpha$ of 6$\times10^{-15}$ cm$^{-2}$ from 
\citet{weg98}, assuming that all transitions are equally probable. 
The average ACE O$^{7+}$ ion flux of $1.0\times10^{5}$ cm$^{-2}$ s$^{-1}$ 
was used as $P_{\rm SWCX}$. Then, the observed line flux $P_{\rm SWCX}$ of 
$\sim$34 LU provided $N_{\rm H}$ of $7\times10^{11}$ cm$^{-2}$. 

As a consistency check, we estimated an average $N_{\rm H}$ from 
the O\emissiontype{VIII} line in the same way. We used $\alpha$ 
of 4$\times10^{-15}$ cm$^{-2}$ based on \citet{bod07}, the O$^{8+}$ 
flux of $2.0\times10^{4}$ cm$^{-2}$ s$^{-1}$, and the observed 
O\emissiontype{VIII} flux of $\sim$13 LU.
The obtained $N_{\rm H}$ was $2\times10^{12}$ cm$^{-2}$,
a factor of 3 larger than the value estimated for the O\emissiontype{VII} line.
This discrepancy may be explained by uncertainties in the assumed parameters such as 
the solar wind oxygen ion fluxes measured with ACE and/or $\alpha$. For instance, if 
the solar wind velocity in the reaction region is lower (e.g., 200 km s$^{-1}$) than 
what we assumed, $\alpha$ for the O\emissiontype{VIII} line can decrease down to 
2.7$\times10^{-15}$ cm$^{-2}$ and $N_{\rm H}$ will increase by a factor of 1.5. 
Since $N_{\rm H}$ estimated from the O\emissiontype{VII} line will decrease only 
by 10\% in this case, a part of the discrepancy can be accounted for.  
If this is the case, the high line flux ratio to the O\emissiontype{VIII} 
in this observation (figure \ref{fig:ratio}) may originate from a relatively 
low O$^{8+}$ ion flux compared to that in \citet{car10}.

For comparison, we estimated the neutral density of hydrogen atoms in the 
Earth's exosphere.
We used the \citet{ost03} model $n(r)$ for the hydrogen density profile around 
Earth and limits this to a minimum density of 0.4 cm$^{-3}$ \citep{fah71}.
$N_{\rm H}$ can be deduced from the integration of the density in the line
of sight as below,
\begin{equation} 
N_{\rm H} = \int_{r_{\rm start}}^{r_{\rm end}}~ n(r)~ dr,
\end{equation}
where $r_{\rm start}$ is the distance to the nearest point 
in the line of sight where the solar wind interacts with the exosphere
and $r_{\rm end}$ is the end point of the integration. 
We roughly assumed $r_{\rm start}$ and $r_{\rm end}$ of 6 and 200 $R_{\rm E}$, 
considering the line of sight geometry (figure \ref{fig:los}).
Here we calculated the geocentric distance of the point where the 
geomagnetic field becomes open to space for the first time in
the line of sight in the same way as \citet{fuj07} and defined
an average distance during the observation as $r_{\rm start}$.
That is, the region should mainly correspond to the magnetosheath.
Then, $N_{\rm H}$ became 2$\times10^{11}$ cm$^{-2}$. 
Hence, this empirical estimation is $\sim$3 and 10 times smaller than the 
observation-based values for O\emissiontype{VII} and O\emissiontype{VIII}, 
respectively.

One possibility for this inconsistency is the neutral hydrogen density
model itself. The hydrogen density may be higher than the model.
Another is that more solar wind ions enter into the magnetosphere
than we assumed, i.e., $r_{\rm start}<6~R_{\rm E}$. In this case, the SWCX reaction 
occurs near Earth where the exospheric density is high, leading to 
a larger theoretical $N_{\rm H}$.

The latter hypothesis might be supported by other observational facts. 
Firstly, the X-ray enhancement in figures \ref{fig:cur1} and \ref{fig:cur2}
coincides with a peak of the geomagnetic storm, corresponding to the duration 
when many solar wind ions entered into the Earth's inner magnetosphere and 
caused disturbed ring currents. Secondly, the cross correlation between the 
high resolution WIND proton data and Suzaku light curve (figure \ref{fig:cross2})
suggests that a positive delay exists even after considering the satellite 
positions. Thirdly, the exospheric SWCX in the magnetosheath direction 
found by \citet{ezo10} was an order of magnitude brighter than the model
calculation assuming that the SWCX occurs outside the magnetopause. 
The invasion of the solar wind inside the magnetopause and resultant 
SWCX reaction can explain the inconsistency as well.
\citet{ebi09} suggested that the O$^{6+}$ ions are transported to the inner 
magnetosphere from the high-latitude magnetopause in August - September 1998 storms
driven by CMEs. 
This result seems to be consistent with the suggestion from the Suzaku observations; 
the O$^{n+}$ ions may enter into the high-latitude magnetopause through the 
magnetosheath region.
To verify this hypothesis, we need an accurate calculation of solar wind 
transportation and SWCX reaction and ideally should conduct a direct imaging 
of the exospheric SWCX at a distant position from Earth in future space missions.

\section{Summary and Conclusion}
\label{sec:summary}

In this paper, we have investigated terrestrial diffuse X-ray emission
associated with the strong geomagnetic storm driven by CMEs on 2005 August
23-24 using Suzaku data. This geomagnetic storm was the strongest in 
terms of the Dst index after the Suzaku launch, as of 2011 April.
We found that this event was a textbook case and provided us with a very 
good opportunity to study the scattering of solar X-rays and the exospheric SWCX. 
We have obtained following results.

\begin{itemize}

\item The X-ray light curve in 0.2--1 keV showed a factor of 2 enhancement,
while that in 2--5 keV was almost constant. The enhancement coincided
with the geomagnetic storm and increase of solar wind proton and O$^{7+}$ fluxes.

\item The X-ray spectrum in the pre-storm period was explained by the sum
of diffuse sky background and leaked photons from bright X-ray sources
within the field of view. 

\item The soft X-ray enhancement in the storm period consisted of two
types of terrestrial diffuse X-ray emission. One is the fluorescent 
scattering of solar X-rays by Earth's low altitude atmosphere seen 
in low ELV angles ($5\sim10$ deg). 
The other is the exospheric SWCX from highly ionized carbon, nitrogen, 
and oxygen. After the removal of the scattering component, the X-ray 
spectrum was well represented by the Bodewits's SWCX model. 

\item A cross correlation analysis was conducted between the 
SWCX O\emissiontype{VII} line and the low time-resolution 
solar wind O$^{7+}$ data. The estimated time delay of 
$<2$ hr was consistent with the expected travelling time 
of the solar wind (1 hr).
The relation between the O\emissiontype{VII} count rate and
solar wind O$^{7+}$ flux showed a possible positive 
correlation.

\item 
The cross correlation between the O\emissiontype{VII} count rate and 
the high time resolution solar wind proton flux suggests the existence 
of an extra delay in addition to the 1 hr delay due to the 
solar wind travelling time from the satellite position to Earth.
The relation between the O\emissiontype{VII} count rate 
and the solar wind proton flux was well represented by a 
linear function. A positive offset in the linear relation 
suggests that, even if the 
solar wind flux is zero, there remains a certain amount of X-ray 
flux due to the sky backgrounds. 

\item From the observed neutral oxygen line intensity, 
the column density of oxygen atoms and molecules $N_{\rm O+O2}$
was estimated as $<2\times10^{15}$ and $4\times10^{16}$ cm$^{-2}$ 
in the pre-storm and storm periods, respectively. These values are 
compared with the theoretical NRLMSIS2000 model. 
$N_{\rm O+O2}$ estimated by the two methods can be consistent with 
each other considering uncertainties, although a factor of $>20$
increase was not consistent with the model and hence will be left
as a future issue for the model.

\item 
Similarly, the column density of neutral hydrogen atoms 
for the exospheric SWCX was estimated as $7\sim20\times10^{11}$ 
cm$^{-2}$ from the observed O\emissiontype{VII} and O\emissiontype{VIII}
line fluxes, while the \citet{ost03} model predicts $2\times10^{11}$ 
cm$^{-2}$. Hence, the calculation may underestimate the neutral 
column density by a factor of 3$\sim$10.

\item 
The small column density based on the exospheric density model suggests 
that our assumption on the SWCX reaction region is not proper and a part 
of solar wind may enter inside the magnetosphere where the hydrogen 
density is high.

\end{itemize}

In conclusion, we have demonstrated that the fluorescent scattering 
and exospheric SWCX in combination with the solar wind data are quite 
useful to examine the Earth's atmospheric density models. 
Following to \citet{ezo10}, we conducted the cross correlation 
analysis and checked the relationship between the X-ray line and 
solar wind ion flux. This set of analyses must help to study the 
exospheric density and can provide a global view 
of the transportation of heavy ion originating from the Sun into 
geospace, which may contribute to the evolution of a 
geomagnetic storm.

The authors thank and acknowledge the WIND, ACE, GOES instrument 
groups for making their data freely available for scientific use. 
Furthermore, we gratefully acknowledge the access to Dst and SYM-H
indices from the World Data Center for Geomagnetism, Kyoto, Japan.

\clearpage


\clearpage

\begin{table}[p]
  \caption{Result of the spectral fit to the pre-storm period shown
  in figure \ref{fig:spec2}.}
  \label{tbl:spec2}

  \begin{center}
    \begin{tabular}{lccccccc}

      \hline
      Component & $N_{\rm H}$$^{\rm a}$ 
      & $\Gamma$/$kT$/$E_c$$^{\rm b}$ 
      & Normalization
      & $f_{\rm X}^{\rm c}$ 
      \\ \hline
      
      Power law$^{\rm d}$  & 9.5 (fixed)
      & 2.0 (fixed)
      & 2.2$\pm{0.1}$$\times10^{2}$
      & 1.4$\times$10$^{-12}$ \\

      NEI$^{\rm e}$  & 0.59 (fixed)
      & 0.45$_{-0.03}^{+0.06}$
      & 9.6$_{-2.1}^{+1.5}$$\times10^{2}$ 
      & 3.6$\times$10$^{-13}$ \\

      LHB (Raymond-Smith)$^{\rm f}$ & 0 (fixed)
      & 0.1 (fixed)
      & 15$\pm3$
      & 7.1$\times$10$^{-14}$ \\

      CXB (Power-law)$^{\rm d}$ & 1.4 (fixed)
      & 1.5 (fixed)
      & 10.4 (fixed)
      & 9.4$\times$10$^{-14}$ (fixed) \\

      Gaussian$^{\rm g}$ & 0 (fixed)
      & 0.53 (fixed)
      & 0.76($<2.4$)
      & 2.2$\times$10$^{-15}$ \\

      \hline
      $\chi^2$/d.o.f. & 569.3/405\\     
 
        \hline
    \end{tabular}
  \end{center}
{\noindent
  $^{\rm a}$ Hydrogen column density in 10$^{21}$ cm$^{-2}$.\\
  $^{\rm b}$ Photon index, plasma temperature in keV, or center energy.\\
  $^{\rm c}$ 0.2--5 keV flux in erg s$^{-1}$ cm$^{-2}$.\\
  $^{\rm d}$ Normalization is in units of photon s$^{-1}$ cm$^{-2}$ str$^{-1}$ keV$^{-1}$ at 1 keV.\\
  $^{\rm e}$ Normalization is in units of 1/$4\pi$ $D^{-2}~(1+z)^{-2}~10^{-14}~\int n_{\rm e}n_{\rm H}~dV$ per steradians, where $D$ is the angular size distance to the source, and $n_{\rm e}$, $n_{\rm H}$ are the electron and hydrogen densities, respectively. The other parameters except for the Fe abundance $Z_{\rm Fe}=0.16_{-0.02}^{+0.03}$ are the same as table 1 in \citet{tam96}.\\  
  $^{\rm f}$ Normalization is in the same units of the NEI component.\\
  $^{\rm g}$ Normalization is in units of photon s$^{-1}$ cm$^{-2}$ str$^{-1}$.

}
\end{table}

\begin{table}[p]
  \caption{Result of the single narrow Gaussian fit to the spectrum 
  shown in figure \ref{fig:spec4}$^{\rm a}$. }
  \label{tbl:spec4}

  \begin{center}
    \begin{tabular}{lccccccc}

      \hline\hline

      $E_c$ & Normalization & $f_{\rm X}$ & Line identification & $\chi^2$/d.o.f. \\ \hline

      525$^{+9}_{-12}$ eV 
      & 14$\pm5$
      & 4.2$\times$10$^{-14}$
      & O\emissiontype{I} (524 eV)
      & 6.1/38\\
 
        \hline
    \end{tabular}
  \end{center}
{\noindent
  $^{\rm a}$ $E_c$ is the line center energy. Normalization is
  in units of photons s$^{-1}$ cm$^{-2}$ str$^{-1}$.
  $f_{\rm X}$ is the energy flux in erg s$^{-1}$ cm$^{-2}$.
  The line width is fixed at 0 eV. \\

}
\end{table}

\begin{table}[p]
  \caption{Result of the seven Gaussian fit to the spectrum 
  shown in figure \ref{fig:spec5}$^{\rm a}$. }
  \label{tbl:spec5}

  \begin{center}
    \begin{tabular}{lccccccc}

      \hline\hline

      Model & $E_c$  & Normalization & $f_{\rm X}$ 
      & Principal line 
      \\ \hline

      1     
      & 248$\pm4$ eV 
      & 32$\pm$8
      & 4.1$\times$10$^{-14}$ 
      & C band lines \\

      2     
      & 304$^{+2}_{-5}$ eV 
      & 200$\pm50$
      & 3.4$\times$10$^{-13}$ 
      & C\emissiontype{V} (299 eV) \\

      3     
      & 372$^{+10}_{-6}$ eV 
      & 40$^{+6}_{-9}$
      & 8.5$\times$10$^{-14}$ 
      & C\emissiontype{VI} 2p to 1s (367 eV) \\

      4     
      & 453$^{+5}_{-10}$ eV 
      & 13$\pm3$
      & 3.5$\times$10$^{-14}$ 
      & C\emissiontype{VI} 4p to 1s (459 eV)\\

      5     
      & 513$^{+5}_{-12}$ eV 
      & 11$\pm4$
      & 3.3$\times$10$^{-14}$ 
      & N\emissiontype{VII} 2p to 1s (500 eV)\\

      6     
      & 565$^{+10}_{-1}$ eV 
      & 34$\pm5$
      & 1.1$\times$10$^{-13}$ 
      & O\emissiontype{VII} (f 561 eV, r 575 eV)$^{\rm b}$ \\

      7     
      & 650$^{+7}_{-13}$ eV 
      & 13$\pm2$
      & 4.8$\times$10$^{-14}$ 
      & O\emissiontype{VIII} 2p to 1s (653 eV) \\

\hline
      $\chi^2$/d.o.f. &  89.9/47 \\
 
        \hline
    \end{tabular}
  \end{center}
{\noindent
  $^{\rm a}$ Definitions of parameters are the same as in table \ref{tbl:spec4}. 
  All the line widths are fixed at 0 eV.\\
  $^{\rm b}$ f and r denote forbidden and resonance lines. 
}
\end{table}

\begin{table}[p]
  \caption{Result of the SWCX model fit to the spectrum 
  shown in figure \ref{fig:spec5}$^{\rm a}$. }
  \label{tbl:spec5-2}

  \begin{center}
    \begin{tabular}{lccccccc}

      \hline\hline

      Ion & Principal energy (eV) & Normalization & $f_{\rm X}$ 
      \\ \hline

      C band lines     
      & 244$\pm6$ eV 
      & 35$_{-7}^{+20}$
      & 4.9$\times$10$^{-14}$ \\

      C\emissiontype{V}     
      & 299
      & 200$_{-50}^{+110}$
      & 3.4$\times$10$^{-13}$ \\

      C\emissiontype{VI}     
      & 367
      & 32$_{-8}^{+14}$
      & 7.4$\times$10$^{-14}$ \\

      N\emissiontype{VI}     
      & 420
      & 7.7$\pm3.8$
      & 1.9$\times$10$^{-14}$ \\
 
      N\emissiontype{VII}     
      & 500
      & 16$\pm4$
      & 5.0$\times$10$^{-14}$ \\

      O\emissiontype{VII}     
      & 561
      & 36$\pm5$
      & 1.1$\times$10$^{-13}$ \\
 
      O\emissiontype{VIII}     
      & 653
      & 12$\pm3$
      & 4.7$\times$10$^{-14}$ \\

\hline
      $\chi^2$/d.o.f. &  77.6/53 \\

        \hline
    \end{tabular}
  \end{center}
{\noindent
  $^{\rm a}$ Definitions of parameters are the same as in table \ref{tbl:spec4}. 
  Only the principle transitions from C, N, and O plus are listed except for
  the low energy C band complex. All the line widths are fixed at 0 eV.\\
}
\end{table}

\end{document}